\documentclass[aps,prx,10pt,twocolumn,superscriptaddress,notitlepage,floatfix]{revtex4-1}

\usepackage[utf8]{inputenc}
\usepackage{graphicx}
\graphicspath{{../}}

\pdfoutput=1 

\usepackage[ruled,vlined]{algorithm2e}

\usepackage[colorlinks=true, hyperindex, breaklinks, linkcolor=blue, urlcolor=blue, citecolor=blue]{hyperref} 

\usepackage{dsfont}
\usepackage{amsmath}
\usepackage{amssymb,bm}
\usepackage{amsthm}
\usepackage{array, makecell}
\usepackage{boldline}
\usepackage[capitalise]{cleveref}
\usepackage{graphicx}
\usepackage[caption=false]{subfig}
\usepackage[normalem]{ulem}
\usepackage{cancel}

\begin{document}

\title{Stabilizing a Bosonic Qubit using Colored Dissipation}

\author{Harald Putterman} 
\email{putterma@amazon.com}
\affiliation{AWS Center for Quantum Computing, Pasadena, CA 91125, USA}
\affiliation{IQIM, California Institute of Technology, Pasadena, CA 91125, USA}

\author{Joseph Iverson} 
\thanks{These authors contributed equally.}
\affiliation{AWS Center for Quantum Computing, Pasadena, CA 91125, USA}
\affiliation{IQIM, California Institute of Technology, Pasadena, CA 91125, USA}

\author{Qian Xu} 
\thanks{These authors contributed equally.}
\affiliation{AWS Center for Quantum Computing, Pasadena, CA 91125, USA}
\affiliation{Pritzker School of Molecular Engineering, The University of Chicago, Illinois 60637, USA}

\author{Liang Jiang} 
\affiliation{AWS Center for Quantum Computing, Pasadena, CA 91125, USA}
\affiliation{Pritzker School of Molecular Engineering, The University of Chicago, Illinois 60637, USA}

\author{Oskar Painter} 
\affiliation{AWS Center for Quantum Computing, Pasadena, CA 91125, USA}
\affiliation{IQIM, California Institute of Technology, Pasadena, CA 91125, USA}

\author{Fernando G.S.L. Brand\~ao} 
\affiliation{AWS Center for Quantum Computing, Pasadena, CA 91125, USA}
\affiliation{IQIM, California Institute of Technology, Pasadena, CA 91125, USA}

\author{Kyungjoo Noh} 
\email{nkyungjo@amazon.com}
\affiliation{AWS Center for Quantum Computing, Pasadena, CA 91125, USA}
\affiliation{IQIM, California Institute of Technology, Pasadena, CA 91125, USA}

\date{\today}

\begin{abstract}
 
Protected qubits such as the 0-$\pi$ qubit, and bosonic qubits including cat qubits and GKP qubits offer advantages for fault-tolerance. Some of these protected qubits (e.g., 0-$\pi$ qubit and Kerr cat qubit) are stabilized by Hamiltonians which have (near-)degenerate ground state manifolds with large energy-gaps to the excited state manifolds. Without dissipative stabilization mechanisms the performance of such energy-gap-protected qubits can be limited by leakage to excited states. Here, we propose a scheme for dissipatively stabilizing an energy-gap-protected qubit using colored (i.e., frequency-selective) dissipation without inducing errors in the ground state manifold. Concretely we apply our colored dissipation technique to Kerr cat qubits and propose colored Kerr cat qubits which are protected by an engineered colored single-photon loss. When applied to the Kerr cat qubits our scheme significantly suppresses leakage-induced bit-flip errors (which we show are a limiting error mechanism) while only using linear interactions.  Beyond the benefits to the Kerr cat qubit we also show that our frequency-selective loss technique can be applied to a broader class of protected qubits.

\end{abstract}

\maketitle

\textit{Introduction--}One standard approach for realizing fault-tolerant quantum computation is to use the surface code \cite{Bravyi1998_quantum} (or its similar variants) with two-level systems such as transmons \cite{Koch2007_charge_insensitive,Schreier2008_suppressing} or trapped-ion qubits \cite{Egan2020_fault_tolerant}. One promising alternative approach is based on protected qubits \cite{GyenisNoiseProtected}. Examples of protected qubits include the 0-$\pi$ qubit \cite{Kitaev2006_protected,Brooks2013_protected,Groszkowski_2018,Gyenis2021_experimental}, and bosonic qubits \cite{Joshi2020_quantum,cai2021bosonic} such as cat qubits \cite{Mirrahimi_2014, Leghtas853} and GKP qubits \cite{GottesmanEncodingQubit}. Such protected qubits can have an intrinsic robustness against environmental errors because of the structure of the wavefunctions and/or vanishing energy dispersion. This physical level of error-suppression can reduce the hardware overhead for implementing fault-tolerance techniques with protected qubits \cite{Guillaud2019Repetition,chamberland2020building}. 

Certain protected qubits such as the 0-$\pi$ qubits and Kerr cat qubits (a type of bosonic qubit) are stabilized by a Hamiltonian. In this case of Hamiltonian protected qubits, the computational basis states are given by (near-)degenerate ground states of a Hamiltonian with an energy gap to the excited state manifolds. Importantly in the absence of dissipative stabilization mechanisms, leaked population in the excited state manifolds (e.g., from incoherent heating) cannot be returned to the code manifold. As we will illustrate below, depending on the structure of the excited states, such leakage can severely limit the performance of energy-gap-protected qubits.

In this Letter, we present a solution to this problem by adding colored (or frequency-selective) dissipation to energy-gap-protected qubits. In particular, we show that colored dissipation with a suitably engineered bath spectrum can bring the excited states back to the code manifold while not causing logical errors in the code manifold. To make the discussion concrete, we first focus on Kerr cat qubits and show how they are limited by leakage-induced bit-flip errors. We then propose \textit{colored Kerr cat qubits}, i.e., Kerr cat qubits that are protected by colored single-photon loss. Specifically, we propose to engineer the bath spectrum of the colored loss channel by using multiple filter modes. See \cref{fig:schematic diagram for the frequency selective loss} for a schematic diagram. We then provide a general formulation of our colored dissipation technique and explain how it can be applied to a wide class of energy-gap-protected qubits besides Kerr cat qubits. 

\begin{figure}
    \centering
    \includegraphics[width=\columnwidth]{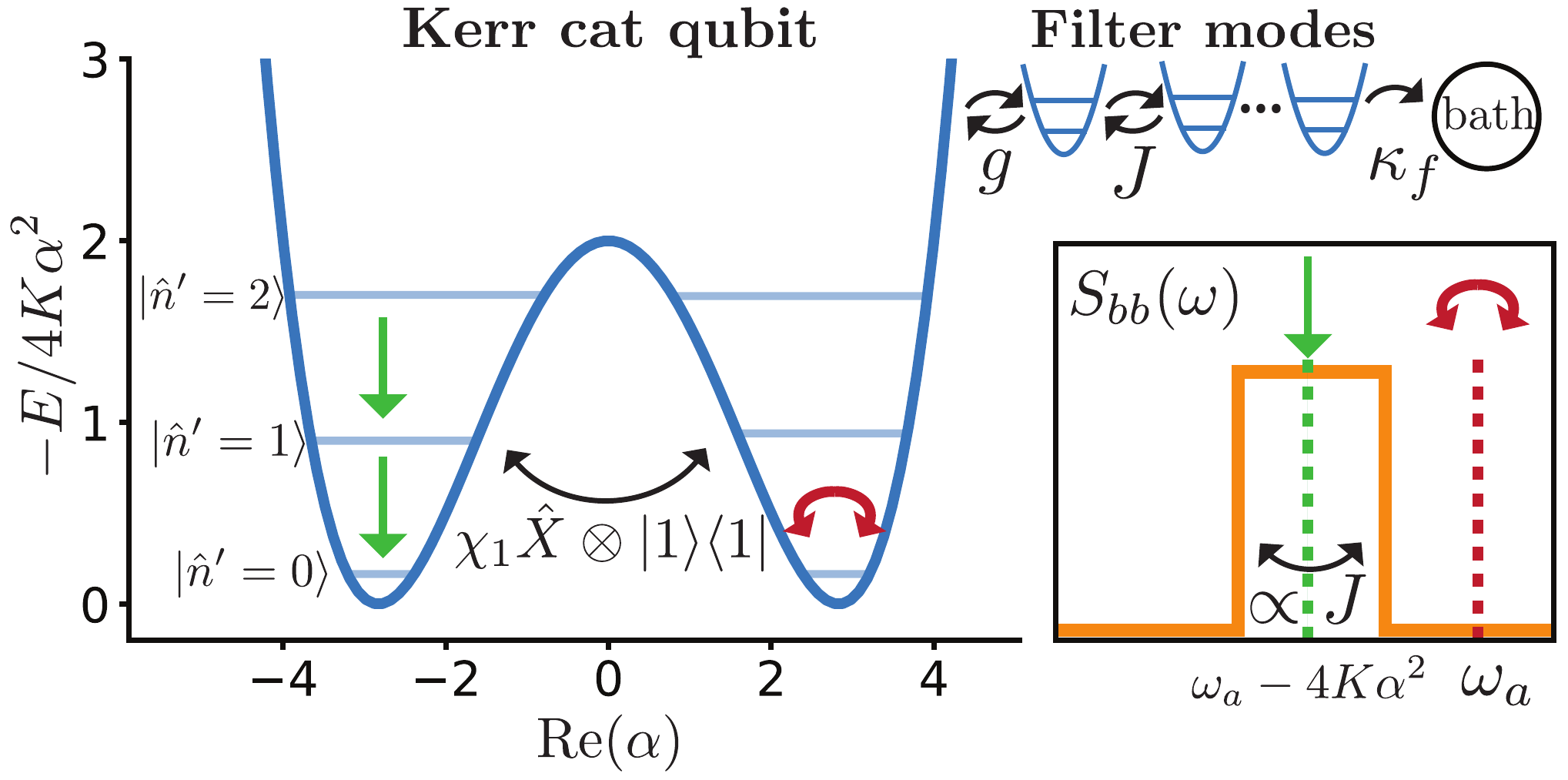}
    \caption{Schematic representation of a colored Kerr cat qubit, i.e., a Kerr cat qubit protected by frequency-selective (colored) single-photon loss. As shown in the inset, the harmonic filter modes are designed such that the colored single-photon loss realizes the desired cooling process (green arrow) while not inducing any additional phase-flip errors in the cat qubit manifold (red arrow). The black arrow in the schematic energy diagram represents the tunneling process between the first excited states, which is important for understanding the bit-flip rate of a Kerr cat qubit.  
    }
    \label{fig:schematic diagram for the frequency selective loss}
\end{figure}

\textit{Cat qubits--} Before proceeding, we briefly summarize the idea behind cat qubits. Two-component cat qubits \cite{cochrane1999macroscopically,jeong2002efficient,Mirrahimi_2014,Leghtas853,Touzard2018_coherent,Lescanne2020,Goto2016_bifurcation,Goto2016_universal,Puri2017,Grimm2020} encode information into an oscillator mode using the $|\pm \alpha\rangle$ coherent states as their approximate computational basis states.  These qubits benefit from an exponential suppression of bit-flip errors with $|\alpha|^2$ due to the large phase space separation between $|\pm\alpha\rangle$ with $|\alpha|^{2} \gg 1$.  This bias against bit-flip (X) errors can be maintained during the execution of gates \cite{Guillaud2019Repetition,Puri2020_bias} allowing us to focus on correcting the dominant phase-flip (Z) errors. This can reduce the hardware overhead of error correction compared to that of bare two-level qubits \cite{Tuckett2018_ultrahigh,Tuckett2019_tailoring,Guillaud2019Repetition,Tuckett2020_fault,Puri2020_bias,chamberland2020building,Guillaud2021_error,BonillaAtaides2021_XZZX,darmawan2021practical}.

When studying cat qubits, we make use of the shifted-Fock basis \cite{chamberland2020building}; a subsystem decomposition which breaks the Hilbert space of our Harmonic modes into two sectors.  The sectors capture the encoded logical information and gauge information about the system.  The shifted Fock basis states are spanned by the displaced Fock states $\hat{D}(\pm \alpha)|\hat{n} = n\rangle$.  As shown in Ref.\ \cite{chamberland2020building},we can express the annihilation operator in this basis as follows.
\begin{align}
    \hat{a} = \hat{Z} \otimes (\hat{a}' + \alpha) + \mathcal{O}( e^{-2|\alpha|^{2}} ). 
    \label{eq:shifted_fock_annihilation_operator}
\end{align}
In this subsystem decomposition, $\hat{Z}$ is a $2\times 2$ Pauli Z operator acting on a qubit sector which describes the cat qubit logical information. The qubit sector of $\hat{a}$ is given by $\hat{Z}$ because single-photon loss changes the parity of the cat qubit and hence causes a phase-flip (or Z) error on the logical information in our basis convention.  $\hat{a}'$ is an annihilation operator acting on a gauge sector which lowers the cat qubit to the ground state manifold (where $\hat{a}'^\dagger\hat{a}'=0$). In what follows, we assume that $\alpha$ is real. 

\textit{Kerr cat qubits}--Kerr cat qubits are an implementation of two-component cat codes that stabilize the $|\pm\alpha\rangle$ manifold using a Hamiltonian with a Kerr nonlinearity and two-photon drive $\hat{H}_{\mathrm{KC}} = -K(\hat{a}^{\dagger 2}-\alpha^2)(\hat{a}^{2}-\alpha^2) $. Rewriting this Hamiltonian in the shifted Fock basis, we find 
\begin{align}
\hat{H}_{\mathrm{KC}} &= -4K\alpha^2 \hat{I} \otimes  \hat{a}'^\dagger \hat{a}' - K\hat{I}\otimes \hat{a}'^{\dagger 2}\hat{a}'^2
\nonumber\\
& \!\!\!\! - 2K\alpha\hat{I}\otimes (\hat{a}'^{\dagger 2}\hat{a}' +  \hat{a}'^\dagger \hat{a}'^2) + \mathcal{O}( e^{-2\alpha^{2}} ). \label{eq:Kerr cat Hamiltonian}
\end{align}  
In the limit of small excitations in the gauge sector (i.e., $\langle \hat{a}'^{\dagger}\hat{a}' \rangle \ll \alpha $), all but the first term in \cref{eq:Kerr cat Hamiltonian} can be neglected and the Kerr cat Hamiltonian is approximately reduced to that of a harmonic oscillator with an energy spacing $-4K\alpha^2$. This non-zero energy gap protects Kerr cat qubits against coherent perturbations by making them off-resonant. However, Kerr cat qubits are not robust against some incoherent perturbations (e.g., heating) due to the absence of a dissipative stabilization mechanism. 

\textit{Heating-induced bit-flip errors--}Heating of an oscillator can be modeled by the dissipator $\kappa_{1}n_{\mathrm{th}}\mathcal{D}[\hat{a}^{\dagger}]$. Since the creation operator $\hat{a}^{\dagger}$ is approximately given by $\hat{a}^{\dagger} \simeq \hat{Z}\otimes (\hat{a}'^{\dagger}+\alpha)$ in the shifted-Fock basis, heating induces phase flips and importantly leakage outside the code space due to the $\hat{a}'^{\dagger}$ term in the gauge sector (see \cref{fig:bit flip numerics} (a)) \footnote{Dephasing $\kappa_{\phi}\mathcal{D}[\hat{a}^{\dagger}\hat{a}]$ can also lead to leakage but the energy gap suppresses $1/f$ noise.  Furthermore dephasing and heating can be treated similarly so we focus on heating \cite{Grimm2020} (see also \cite{supplement}).}.  

Indeed, in the first experimental realization of a Kerr cat qubit \cite{Grimm2020}, significant heating occurred and only a modest noise bias factor of $\sim 40$ was achieved.  Thus, realizing the full potential of Kerr cat qubits requires counteracting the leakage caused by heating.

To be used as a biased-noise qubit, Kerr cat qubits need to have strongly suppressed bit-flip errors with $\alpha^2$. Previous works \cite{Puri2019Stabilized,Grimm2020} have suggested that Kerr cat qubits can be made robust to leakage induced bit-flip errors by ensuring that higher excited states reached through heating are below the energy barrier so that that tunneling \cite{Marthaler2007_quantum, Lin2015_critical} between them is suppressed. Although this argument is qualitatively correct, we show that it does not apply to near term experiments and fault-tolerant quantum computation proposals where heating poses a limit on achievable bit-flip times in both regimes.

In \cref{fig:bit flip numerics}, we consider a set of experimentally relevant parameters: $K = 2\pi \times 10$MHz, $\kappa_{1} = 2\pi \times 1$kHz (corresponding to the lifetime of $1/\kappa_{1} = 159\mu s$), and a thermal populations of  $n_{\mathrm{th}} = 0.1$ and $n_{\mathrm{th}} = 0.01$ \footnote{The chosen values of $K$ and $n_{\mathrm{th}}=.1$ are close to those in \cite{Grimm2020} while the lifetime is 10 times larger so it is in a known regime for fault tolerant quantum computation \cite{darmawan2021practical}}. As indicated by the blue line in \cref{fig:bit flip numerics} (b), the bit-flip error rate $\gamma_{X}$ of a Kerr cat qubit stays constant throughout the range $3 \le \alpha^{2} \lesssim 9$, which are most experimentally relevant. This contrasts with expectations for exponential suppression of the bit-flip error rate $\gamma_{X}$ with $\alpha^{2}$ used throughout the literature for biased noise cat qubits.    

To understand why the bit-flip error rate $\gamma_{X}$ of a Kerr cat qubit does not improve as we increase $\alpha^{2}$ up to $9$, we need to consider the $\mathcal{O}(e^{-2\alpha^{2}})$ contributions in \cref{eq:Kerr cat Hamiltonian}. In particular, we need to consider the terms in $\hat{H}_{\mathrm{KC}}$ of the form $\chi_{n}\hat{X}\otimes |\hat{n}' = n\rangle\langle \hat{n}' = n|$.  Here, $\chi_{n}$ can be understood as the tunneling rate between the states $|0\rangle\otimes |\hat{n}'=n\rangle$ and $|1\rangle\otimes |\hat{n}'=n\rangle$ (see the schematic \cref{fig:schematic diagram for the frequency selective loss}). In \cite{supplement}, we show that the tunneling rate $\chi_{1}$ in the first excited state manifold is perturbatively given by 
\begin{align}\label{eq:chi_one}
    \chi_{1} \simeq 16K\alpha^{4} e^{-2\alpha^{2}} , 
\end{align}
which agrees with the exact numerical results for all $\alpha^{2} \ge 3$. Although $\chi_{1}$ decreases exponentially in $\alpha^{2}$, the large prefactor $16K\alpha^{4}$ can still make this $\chi_{1}$ (induced by the Kerr cat Hamiltonian $\hat{H}_{\mathrm{KC}}$) limiting in practice.  

We now explain why the bit-flip error rate $\gamma_{X}$ (blue line in \cref{fig:bit flip numerics} (b)) plateaus in the range $3\le \alpha^{2} \lesssim 9$.  Heating excites the system to the first excited state manifold.  Here it persists for a time $\Delta t \sim 1/\kappa_{1}$ until it decays back to the cat state manifold. During this period, if $\chi_1 \gg \kappa_1$, rapid oscillations occur between the states $|0\rangle\otimes |\hat{n}' = 1\rangle$ and $|1\rangle\otimes |\hat{n}' = 1\rangle$. In this regime, a bit-flip error happens with $50\%$ probability whenever heating creates an excitation. As a result, the bit-flip error rate is given by half the heating rate, i.e., $\gamma_{X} = \kappa_{1}n_{\mathrm{th}}/2$ in the regime of $\chi_{1}\gg \kappa_{1}$. With our parameters (yielding $K/\kappa_{1} = 10^{4}$), $\chi_{1} = 16K\alpha^{4}e^{-2\alpha^{2}}$ is at least $10$ times larger than $\kappa_{1}$ for all $3 \le \alpha^{2} \le 6.75$ and $\chi_{1} = \kappa_{1}$ at $\alpha^{2} = 8.08$. This explains why the bit-flip error rate $\gamma_{X}$ is independent of $\alpha^{2}$ and given by $\kappa_{1}n_{\mathrm{th}}/2$ in the range $3 \le \alpha^{2} \lesssim 9$. Above $\alpha^2 \sim 9$ heating to higher excited states becomes the important error mechanism because tunneling between the first excited states is sufficiently suppressed (see \cite{supplement}). A similar mechanism can limit other energy-gap-protected qubits if the transition rates within the excited state manifold are significant.

\begin{figure}
    \centering
    \includegraphics[width=.9\columnwidth]{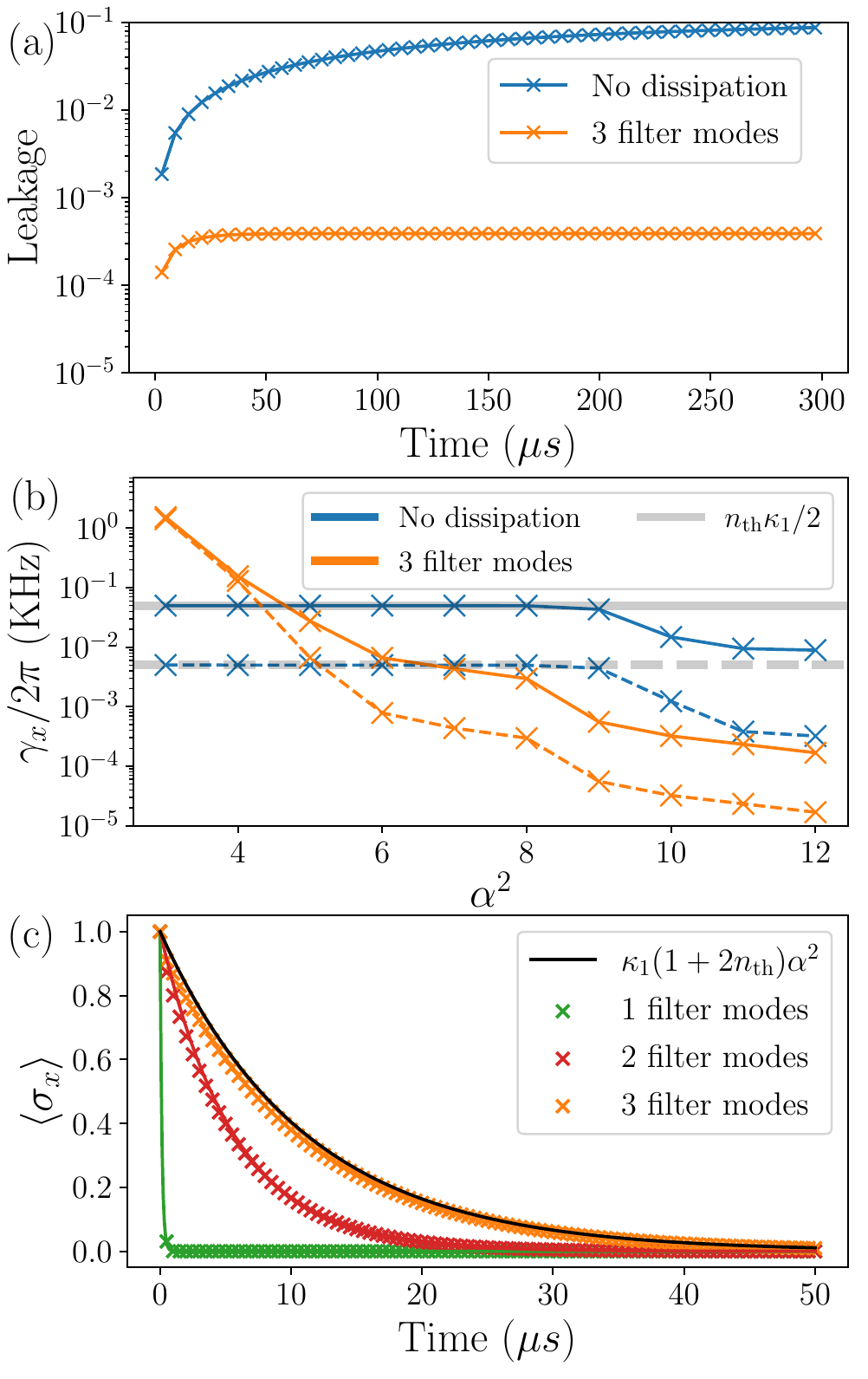}
    \caption{(a) Leakage accumulation over time in a Kerr cat qubit without any engineered dissipation (blue) and a colored Kerr cat qubit with three filter modes (orange) from the initial state $|\alpha\rangle$. (b) Bit-flip error rate of a Kerr cat qubit (blue) and a colored Kerr cat qubit with three filter modes (orange) as a function of the average photon number $\alpha^{2}$ for $n_{\mathrm{th}}=0.1$ (solid) and $0.01$ (dotted). Grey lines represent the analytical prediction $\gamma_{X} = \kappa_{1}n_{\mathrm{th}}/2$ for the regime $\chi_{1} \gg \kappa_{1} + \kappa_{1,\mathrm{eng}}$. (c) Decay of the parity of a colored Kerr cat qubit starting from $|+\rangle$ with one (green), two (red), and three (orange) filter modes. Xs represent numerical data and solid lines represent analytical predictions. The black line shows the baseline decay of the parity at a rate $ 2\kappa_1(1+n_\mathrm{th})\alpha^2 $, i.e., twice the phase-flip rate. In all three plots, we use the parameters $K=2\pi \times 10\mathrm{MHz}$ and $\kappa_{1} = 2\pi \times 1\mathrm{kHz}$. In (a) and (c), we further assume $n_{\mathrm{th}} = 0.1$ and $\alpha^{2}=6$. See \cite{supplement} for  details. 
    }
    \label{fig:bit flip numerics}
\end{figure}
 
\textit{Colored Kerr cat qubits}--As shown by our numerical and analytical results, the heating-induced bit-flip errors can be even more detrimental than previously anticipated. Here, we propose to counteract the heating and leakage by adding frequency-selective (i.e., colored \cite{MarquardtQuantumTheory, MurchCavityAssisted}) single-photon loss to Kerr cat qubits, hence making them \textit{colored Kerr cat qubits}. Our scheme fundamentally differs from the previous proposals based on two-photon dissipation \cite{Puri2019Stabilized,Grimm2020,darmawan2021practical} as we only require single-photon loss. Intrinsic single-photon loss $\kappa_{1}\mathcal{D}[\hat{a}]$ is harmful for cat qubits because the $+\alpha \hat{Z}\otimes \hat{I}$ term in the shifted-Fock basis representation of the annihilation operator $\hat{a} \simeq \hat{Z}\otimes (\hat{a} ' + \alpha)$ causes phase-flip (or Z) errors in their ground state manifold \cite{Mirrahimi_2014,Puri2017}. The other term (i.e., $\hat{Z}\otimes \hat{a}'$) is useful for suppressing leakage as it brings the excited states back to the code space via $\hat{a}'$. 

Our key idea is to engineer the frequency spectrum of the bath of the extrinsic single-photon loss such that we can take advantage of the beneficial decay term ($\hat{Z}\otimes \hat{a}'$) while filtering out the parasitic term ($+\alpha \hat{Z}\otimes \hat{I}$) from the single-photon loss $\hat{a}$. Since only the extrinsic single-photon loss is engineered with this technique, the intrinsic single-photon loss rate $\kappa_{1}$ should still be kept as small as possible.

To demonstrate how our scheme works, we introduce a concrete setup where a Kerr cat qubit is coupled to an engineered bath through a set of harmonic filter modes with nearest-neighbor hopping, forming a colored Kerr cat qubit (diagram in \cref{fig:schematic diagram for the frequency selective loss}). Specifically, we consider the following Lindblad equation in the rotating frame of a Kerr cat qubit ($\hat{a}$ with frequency $\omega_{a}$) and filter modes ($\hat{f}_{1}, ..., \hat{f}_{M}$ with frequency $\omega_{f}$):
\begin{align}
    \frac{d\hat{\rho}(t)}{dt} &= -i[\hat{H}, \hat{\rho}(t)] + \kappa_{1}(1+n_{\mathrm{th}})\hat{D}[\hat{a}]\hat{\rho}(t) 
    \nonumber\\
    &\quad + \kappa_{1}n_{\mathrm{th}}\hat{D}[\hat{a}^{\dagger}]\hat{\rho}(t)  + \kappa_{f}\mathcal{D}[\hat{f}_{M}]\hat{\rho}(t), 
\end{align}
where the Hamiltonian $\hat{H}$ is given by
\begin{align}
    \hat{H} = \hat{H}_{\mathrm{KC}} + \Big{[} g \hat{a}\hat{f}_{1}^{\dagger}e^{i\Delta t} + J\sum_{j=1}^{M-1}\hat{f}_{j}\hat{f}_{j+1}^{\dagger} + \mathrm{h.c.}\Big{]} .
    \label{eq:FilterHamiltonian}
\end{align}
Here, $\Delta \equiv \omega_{f}-\omega_{a}$ is the detuning between the filter modes $\hat{f}_{1},\cdots,\hat{f}_{N}$ and the mode $\hat{a}$ which hosts the Kerr cat qubit. Also, $\mathcal{D}[\hat{A}]\hat{\rho} \equiv \hat{A}\hat{\rho}\hat{A}^{\dagger} - \frac{1}{2}\lbrace \hat{A}^{\dagger}\hat{A}, \hat{\rho} \rbrace$ is the Lindblad dissipator. Besides having the intrinsic loss and heating processes, the Kerr cat qubit can lose an excitation to the first filter mode at a rate $g$. Such an excitation is then transported to the last filter mode at a hopping rate $J$ where it decays to a cold bath at a rate $\kappa_{f}$. It is important that this bath and the filter modes have a temperature much lower than the Kerr cat qubit so as to not induce additional heating (here $n_\mathrm{th, filter}=0$).  In practice, strong pump tones are necessary to realize Kerr cat qubits while the filter modes and their bath are passive and undriven.  Thus it is plausible that the filter modes would be colder than the Kerr cat qubits. We take $\kappa_{f} = 2J$ so that the filter modes act as an ideal band-pass filter (centered at the frequency $\omega_{f}$ and with a bandwidth $4J$) as $N\rightarrow \infty$. See \cite{supplement} for more details.          

Recall that in the shifted Fock basis, the Kerr cat Hamiltonian is approximately given by $\hat{H}_{\mathrm{KC}}\simeq -4K\alpha^{2}\hat{I}\otimes \hat{a}'^{\dagger}\hat{a}'$. Transforming to the shifted-Fock basis, and moving into the rotating frame of the $\hat{a}'$ mode the coupling term $g \hat{a}\hat{f}_{1}^{\dagger}e^{i\Delta t}$ becomes $g (\hat{Z}\otimes \hat{a}') \hat{f}_{1}^{\dagger}e^{i(\Delta + 4K\alpha^{2}) t} + g\alpha (\hat{Z}\otimes\hat{I}) \hat{f}_{1}^{\dagger}e^{i\Delta t}$. The first term realizes a desired cooling effect through $\hat{a}'$ whereas the second term causes undesired phase-flip (Z) errors in the cat qubit manifold. By choosing $\Delta = -4K\alpha^{2}$ (or equivalently $\omega_{f} = \omega_{a} - 4K\alpha^{2}$), we can make the desired first term resonant while making the undesired second term off-resonant. Furthermore, by ensuring that the half bandwidth $\kappa_{f} = 2J$ is smaller than the detuning $|\Delta|$, we can place the undesired second term outside the filter passband and filter it out (see \cref{fig:schematic diagram for the frequency selective loss}). In particular, through adiabatic elimination (see \cite{supplement}), the induced phase-flip error rate due to the second term is given by $(4g^2\alpha^{2}/\kappa_{f})\times (J/\Delta)^{2M}$ in the $\Delta\gg J$ limit and hence decreases exponentially in the number of the filter modes $M$. On the other hand, the resonant desired term realizes an engineered cooling process $\kappa_{1,\mathrm{eng}}\mathcal{D}[ \hat{Z} \otimes \hat{a}' ]$ with an effective cooling rate $\kappa_{1,\mathrm{eng}} = 4g^{2}/\kappa_{f}$.  

In \cref{fig:bit flip numerics}, we study the performance of a bare Kerr cat qubit and colored Kerr cat qubits with varying number of filter modes. For colored Kerr cat qubits, we choose $\kappa_{f} = 2J = \Delta/5$ and $g=\kappa_{f}/5$ to filter out the induced  phase-flip errors and guarantee the validity of the adiabatic elimination, respectively. We tune $\Delta = -3.6K\alpha^{2}$ (vs. $\Delta = -4K\alpha^{2}$) by accounting for higher order contributions to more closely target the $0\leftrightarrow1$ transition of the Kerr excited states \cite{supplement}. With these parameters, we get a large engineered cooling rate of $\kappa_{1,\mathrm{eng}} = 2\pi \times 1.15\alpha^{2}$MHz (e.g., $\kappa_{1,\mathrm{eng}} = 2\pi \times 6.9$MHz at $\alpha^{2}=6$). As indicated by the orange line in \cref{fig:bit flip numerics} (a), the leakage population of a Kerr cat qubit (of size $\alpha^{2}=6$) can be made orders of magnitude smaller by adding a frequency-selective single-photon loss with three filter modes. Additionally, the idling bit-flip error rate is reduced by at least an order of magnitude for all $\alpha^{2}\ge 6$ (see \cref{fig:bit flip numerics} (b)). This is because the large engineered cooling rate dramatically reduces the lifetime of excited states (especially the first excited states) so that the condition $\kappa_1+\kappa_{1,\textrm{eng}}\gg \chi_1$ is satisfied at lower values of $\alpha^2$ \footnote{We explain why the bit-flip error rate of a colored Kerr cat qubit is higher than that of a bare Kerr cat qubit in the $\alpha^{2}\lesssim 5$ regime (compare the blue and orange lines in \cref{fig:bit flip numerics} (b)) in \cite{supplement}}.  

In \cref{fig:bit flip numerics} (c) we show the parity as a function of time with 1, 2, and 3 filter modes and $\alpha^2 =6$. With only one or two filter modes, the induced phase-flip rate is much larger than the intrinsic phase-flip rate of $\approx\kappa_{1}(1+2n_{\mathrm{th}})\alpha^{2}$ (green and red lines). With three filter modes, however, the induced phase-flip rate is negligible and the total phase-flip probability is close to the intrinsic rate (orange line). The simulated (Xs) parity decays are consistent with our analytical prediction (solid lines) on the induced phase-flip rate $(4g^2\alpha^{2}/\kappa_{f})\times (J/\Delta)^{2M}$ in the $\Delta\gg J$ limit. Hence, \cref{fig:bit flip numerics} (c) demonstrates that with a properly engineered single-photon loss spectrum we can  benefit  from  the  desired  cooling  effects without inducing additional phase-flip errors. 

\textit{General formulation}--We now present how our colored dissipation technique can be generally applied to a wide class of energy-gap-protected qubits.  Specifically we consider energy-gap-protected qubits whose Hamiltonian is given by $\hat{H}_{\mathrm{PQ}} = \Delta \hat{I} \otimes |1\rangle\langle 1|$. Here, $|0\rangle$ and $|1\rangle$ in the gauge sector respectively correspond to the ground and first-excited state manifolds of the qubit and $\Delta$ is the energy gap.  The Hamiltonian including the filter is given by $\hat{H} = \hat{H}_{\mathrm{PQ}} + \Big{[} g \hat{c}\hat{f}_{1}^{\dagger}e^{i\Delta t} + J\sum_{j=1}^{M-1}\hat{f}_{j}\hat{f}_{j+1}^{\dagger} + \mathrm{h.c.}\Big{]}$, where $\hat{c}$ is a coupling operator acting on the protected qubit. In the subsystem decomposition this coupling operator generically takes the form $\hat{c}=\sum_{i,j\in \lbrace 0, 1\rbrace} \hat{c}_{ij} \otimes |i\rangle\langle j|$.  The limit of interest is when $|\Delta|\gg J$ such that any induced  dissipation can be selective on  the $|1\rangle \rightarrow |0\rangle$ decay.  Similarly as above, adiabatic elimination of the filter modes  yields a Lindblad term $(4g^{2}/\kappa_{f})\mathcal{D}[ \hat{c}_{01} \otimes |0\rangle\langle 1| ]$ realizing the desired $|1\rangle \rightarrow |0\rangle$ decay. Crucially the incoherent errors induced on the ground state manifold are exponentially suppressed with the number of filter modes. Specifically these errors are described by the Lindblad term  $(4g^2/\kappa_{f})\times (J/\Delta)^{2M}\mathcal{D}[\hat{c}_{00}]$ upon adiabatic elimination of the gauge mode as well as the filter modes.  Thus if one uses sufficiently many filter modes with $|\Delta|\gg J$ and $\hat{c}$ has  an appreciable matrix element for $\hat{c}_{01}$ one can realize dissipative confinement to the ground state manifold without inducing incoherent errors on the logical information even when $\hat{c}_{00}$ is non-trivial and acts as a logical error in the ground state manifold (\footnote{There is also a coherent term to consider $\hat{H}=-\frac{1}{\Delta}\hat{c}_{00}^\dagger \hat{c}_{00}$ related to Kerr-cat cases low $\alpha^2$ scaling \cite{supplement}}).

\textit{Discussion and outlook--}An interesting future direction is to apply our colored dissipation scheme to other energy-gap-protected qubits. An example is the Hamiltonian-stabilized finite-energy GKP qubit where a gap opens up relative to the infinite-energy case \cite{GottesmanEncodingQubit,RoyerFiniteEnergy,Le2019,Rymarz2021,Conrad2021}.  In this case the subsystem decomposition would be given by a finite energy version of the modular bosonic subsystem decomposition \cite{PantaleoniModularBosonic}.  The 0-$\pi$ qubit similarly \cite{Kitaev2006_protected,Brooks2013_protected,Gyenis2021_experimental, Groszkowski_2018} would be an interesting case with near degenerate ground states enabling frequency-selective loss.  The use of colored loss can also extended to the application of gates.  We consider this for Kerr cat qubits in \cite{supplement}.

In practice the optimal choice of the filter may not be a bandpass filter centered around the gap frequency (as in \cref{fig:schematic diagram for the frequency selective loss}). Other filter geometries such as wider bandpass filters with the frequency $\omega-\Delta$ near the edge of the passband or low pass filters may allow for higher dissipation rates while still rejecting signals at $\omega$ and are interesting areas for future work. These filters can be implemented experimentally in superconducting circuits using quantum metamaterials \cite{ferreira2020collapse}.

In the context of Kerr cat qubits our proposal takes advantage of the energy structure of the gauge mode as opposed to the parity symmetry of two-photon dissipation ($\kappa_{2}\mathcal{D}[\hat{a}^{2}-\alpha^{2}]$ \cite{Puri2019Stabilized,Grimm2020,Puri2020_bias,darmawan2021practical}) such that it only requires single-photon loss.  In particular, this means non-linear interactions are not needed to implement the dissipation potentially enabling larger engineered cooling rates.

Additionally unlike two-photon dissipation, our engineered cooling process $\kappa_{1,\mathrm{eng}}\mathcal{D}[\hat{Z}\otimes \hat{a}']$ comes with a phase-flip $\hat{Z}$ in the qubit sector.  This phase-flip is not problematic because it is only triggered in the excited state manifold.  Moreover, some leakage processes such as those associated with heating ($\hat{a}^\dagger$) and the $Z$ gate ($\hat{H}_Z = \epsilon_Z (\hat{a}^{\dagger} + \hat{a})$) come with a phase-flip in the qubit sector.  There having the phase flip in the cooling process is a feature because the phase flip from the leakage process is canceled out when the system is brought back to the ground state manifold via the colored dissipation. Non-adiabatic gate errors can also be directly suppressed by the frequency selectivity of the filter if the leakage processes they are associated with are off-resonance from the filter. We remark that there is a complementary approach for suppressing bit-flip error rates by reducing the effective tunneling rates $\chi_{1},\chi_{2},\cdots$, which can be done by adding a linear drive to the Kerr-cat Hamiltonian.  These interesting areas for future work are discussed in \cite{supplement}. 

\textit{Acknowledgment--}We thank Arne L. Grimsmo, Matthew H. Matheny, and Gil Refael for useful comments on the manuscript. 

\bibliography{Kerr_cat}

\title{Supplementary Material for "Colored Kerr Cat Qubits: Protecting a Bosonic Qubit using Colored Dissipation"}

\maketitle

\clearpage

\renewcommand*\tocname{Supplementary Material}
\tableofcontents

\section{Leakage and bit-flip induced by dephasing}
\label{sec:dephasing_heating}

In addition to the loss and heating, the non-linear oscillator also suffers from dephasing. If the noise source of the dephaisng had a white spectrum, the dynamics of the Kerr-cat qubit would be described by the dissipator $\kappa_{\phi, 01}\mathcal{D}[\hat{a}^{\dagger}\hat{a}]$, where $\kappa_{\phi, 01}$ is the effective dephasing rate which can be obtained by measuring the rate of phase decay of a $\lbrace |0\rangle,|1\rangle \rbrace$ encoded qubit. The rate of heating of the Kerr-cat qubit under this white-dephasing model is given by:
\begin{equation}
    \kappa_{ 0 \rightarrow 1} = \alpha^2 \kappa_{\phi, 01}
\label{eq:white-dephasing}
\end{equation}
where $\kappa_{ 0 \rightarrow 1}$ denotes the rate of heating from the ground states to the first pair of excited states of the Kerr-cat qubit.  As shown in \cite{Grimm2020}, the measured dephasing rate $\kappa_{\phi, 01}$ (using the $\lbrace |0\rangle,|1\rangle \rbrace$ encoded qubit) can be even larger than the photon loss rate $\kappa_1$. As a result, the dephasing induced heating estimated by \cref{eq:white-dephasing} would have been the dominant source of leakage and bit-flip errors for the Kerr-cat qubit. 

However, in experiments \cref{eq:white-dephasing} is an overestimation since the dephasing originates from the common $1/f$ flux noise in superconducting circuits \cite{Grimm2020}. The Hamiltonian of the Kerr-cat qubit under such a noise can be modeled as:
\begin{equation}
\hat{H} = - K(\hat{a}^{2\dagger} - \alpha^2)(\hat{a}^2 - \alpha^2) + f(t) \hat{a}^{\dagger}\hat{a} 
\label{eq:dephasing_f}
\end{equation}
where $f(t)$ is some classical noise with $1/f$-type spectral density. In the shifted Fock basis, \cref{eq:dephasing_f} is approximately given by
\begin{equation}
    \hat{H} \simeq \hat{I} \otimes [ - 4 K\alpha^2  \hat{a}^{\prime \dagger}\hat{a}^{\prime} + \alpha f(t) (\hat{a}^{\prime} + \hat{a}^{\prime \dagger}) + f(t)\hat{a}^{\prime \dagger}\hat{a}^{\prime} ]
\end{equation}
where the second term can induce transitions from ground states to excited states of the Kerr-cat qubit that lead to leakage. The transition rate from 0 to 1 excitation level can be calculated using the Fermi's golden rule:
\begin{equation}
    \kappa_{0\rightarrow 1} = \alpha^2 S_{ff}(-\Delta)
\end{equation}
where $S_{ff}(\omega) \equiv \int_{-\infty}^{\infty} dt \overline{f(t)f(0)}$ is the spectral density of the noise $f(t)$, $\Delta = - 4 K \alpha^2$ is the energy gap of the Kerr-cat qubit. In contrast, the phase-decay rate of a $\lbrace |0\rangle,|1\rangle\rbrace$ encoded qubit with such a colored noise is given by $\kappa_{\phi, 01} = \frac{1}{2}S_{ff}(0)$. Since $S_{ff}(\omega)$ has a $1/f$ spectrum, in contrast to \cref{eq:white-dephasing} we should instead have $ \kappa_{0\rightarrow 1} \ll \alpha^2 \kappa_{\phi, 01}$. 

\section{Calculation of the interwell coupling in the shifted Fock basis}\label{subsection: shifted fock interwell coupling}
Here we compute the coupling rate between the $|0\rangle\otimes| \hat{n}' =1\rangle$ and $|1\rangle\otimes|\hat{n}'=1\rangle$ shifted-Fock excited states  due to the Kerr cat Hamiltonian $\hat{H}_{\mathrm{KC}} = -K(\hat{a}^{\dagger 2}-\alpha^{2})(\hat{a}^{2}-\alpha^{2})$.  The specific term we want to extract from the Kerr cat Hamiltonian is $\chi_1 \hat{X}\otimes|\hat{n}'=1\rangle\langle\hat{n}'=1|$.  In the qubit sector the $\hat{X}$ indicates a coupling which causes tunneling between the two wells of the Kerr cat qubit and in the gauge sector $|\hat{n}'=1\rangle\langle\hat{n}'=1|$ indicates a coupling between the first excited shifted-Fock states.

For this derivation we closely follow Appendix C of \cite{chamberland2020building}. We restrict the cutoff dimension of the shifted-Fock basis to $d_{\max}=2$ since we are only concerned with finding $\chi_1$.  The strategy we follow is to  find the $\mathcal{O}(e^{-2\alpha^2})$ contribution to the lowering operator $\hat{a}$ in the shifted-Fock basis and use this more general form of the lowering operator to find the leading order interwell coupling in the Kerr cat Hamiltonian.

We begin by finding the representation of the lowering operator in the orthonormalized shifted-Fock basis.  To study $Z$ error rates we can directly use the unorthonormalized basis states but in order to study $X$ and $Y$ error rates is necessary to consider the $\mathcal{O}(e^{-2\alpha^2})$ corrections that result from fully orthonormalizing the basis.  Our starting point is the unorthonormalized shifted-Fock basis states
\begin{align}
    |\phi_{n,\pm}\rangle = \frac{1}{\sqrt{2}}\big{[}\hat{D}(\alpha)\pm (-1)^n \hat{D}(-\alpha)\big{]}|\hat{n}=n\rangle.
\end{align}
The basis states are divided into even ($+$) and odd ($-$) branches. The states in the even branch are inherently orthogonal to the states in the odd branch.  The remaining orthonormalization that needs to be done is within the even and odd branches.  

As a first step we define the overlap matrices $\Phi_{m,n}^\pm \equiv \langle \phi_{m,\pm}|\phi_{n,\pm}\rangle$ which describe the overlap between the shifted-Fock states in each parity branch.  With the cutoff dimension $d_{\max}=2$, the overlap matrices are given by
\begin{align}
\Phi_\pm = 
\begin{bmatrix}
1\pm e^{-2\alpha^2} & \mp 2e^{-2\alpha^2} \\
\mp 2e^{-2\alpha^2} & 1\mp e^{-2\alpha^2}(1-4\alpha^2)
\end{bmatrix}.
\end{align}
Next we compute the coefficients $c_{i,j}$ which generate the orthonormal basis set in the even and odd branch of the shifted-Fock basis.  More formally they are defined as part of $|\psi_{n,\pm}\rangle = \sum_{m=0}^{d-1} c_{m,n}^\pm |\phi_{m, \pm}\rangle$.  In terms of the overlap matrices, the orthonormalization coefficients are found to be
\begin{align}
c = \begin{bmatrix}
\frac{1}{\sqrt{\Phi_{0,0}}} & -\frac{\Phi_{0,1}}{\Phi_{0,0}\sqrt{(\Phi_{1,1}-\Phi_{0,1}^2/\Phi_{0,0})}}\\
0 & \frac{1}{\sqrt{(\Phi_{1,1}-\Phi_{0,1}^2/\Phi_{0,0})}}
\end{bmatrix}, 
\end{align}
using the Gram-Schmidt orthonormalization, where the $\pm$ is implicit on all terms.  To the leading order the matrices are 
\begin{align}
c_\pm &\approx \begin{bmatrix}
1 & \pm 2\alpha e^{-2\alpha^2} \\
0 & 1\mp 2\alpha^2 e^{-2\alpha^2}.
\end{bmatrix}.
\end{align}
Now we can compute the form of the lowering operator in the orthonormalized basis by forming the product 
\begin{align}
a\equiv (\hat{H}\otimes\hat{I})c^\dagger \Phi (\hat{X}\otimes(\hat{a}'+\alpha))c(\hat{H}\otimes\hat{I}).
\end{align}
Performing the computation and taking the leading order terms we find that the lowering operator is given by
\begin{align}
    \hat{a} \simeq \hat{Z}\otimes(\hat{a}'+\alpha) -i \hat{Y}\otimes \begin{bmatrix}
    \alpha  & -2\alpha^2  \\
    0 & 4\alpha^3
    \end{bmatrix}e^{-2\alpha^2}, 
    \label{eq:shifted_fock_annihiliation_operator_sup}
\end{align}
in the shifted Fock basis. 

\begin{figure}
    \centering
    \includegraphics[width=\columnwidth]{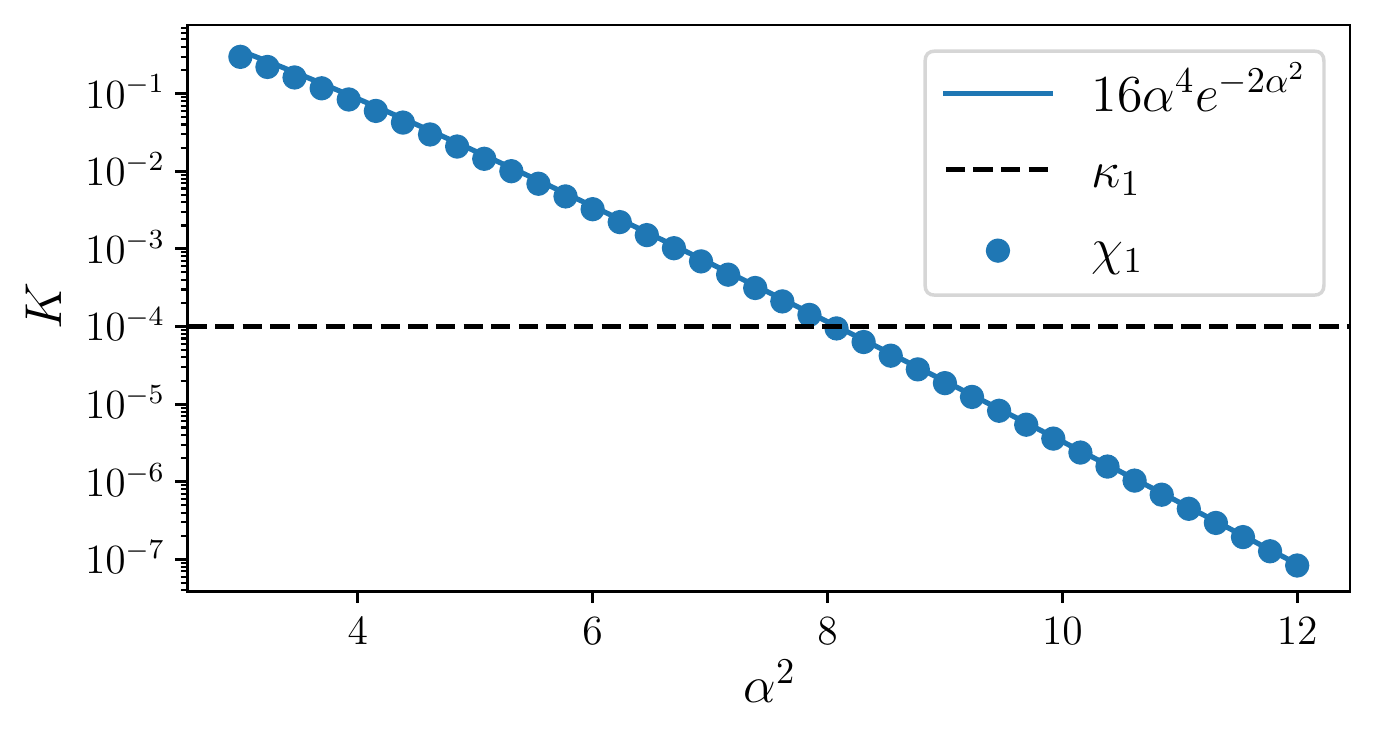}
    \caption{Plot of the interwell coupling of the first shifted-Fock excited state ($\chi_1$) and the interwell coupling of the first and second Kerr excited states ($\chi_{1,2}'$).  The dots indicate numerically extracted coupling rates.   Very good agreement is shown between the numerical $\chi_1$ and the perturbative result \cref{eq:chi_one_sup}  (blue line).  The dashed line corresponds to the lifetime of the Kerr cat qubit which is $\kappa = 10^{-4} K$.}
    \label{fig:interwell_coupling}
\end{figure}

With the lowering operator in hand we can now find the desired term in the Kerr Hamiltonian.  The $\chi_1 \hat{X}\otimes|n'=1\rangle\langle n'=1|$ coupling between the wells of the Kerr cat qubit comes from the Kerr term $-K\hat{a}^{\dagger 2} \hat{a}^2$.  The leading order $\hat{X}$ coupling originates from three $\hat{Z}$ terms and one $\hat{Y}$ term in the 4 different permutations.  Off diagonal couplings and diagonal terms from $\hat{a}^2$ and $\hat{a}^{\dagger 2}$ are subleading.  Substituting and keeping only the leading order term we find that 
\begin{align}
    \chi_1 \simeq 16 K \alpha^4 e^{-2\alpha^2} . 
    \label{eq:chi_one_sup}
\end{align}
In \cref{fig:interwell_coupling} we show good agreement between this analytical prediction and an exact numerical computation.  

\section{Filter Adiabatic Elimination}
\label{sec:filter_elimination}
In this section we give more detail on the adiabatic elimination used to derive the engineered loss rate and induced phase-flip rate.  We closely follow the methods from  \cite{ReiterEffective} and Appendix B of \cite{chamberland2020building}.
\subsection{Engineered Loss}
Here we show how the tight-binding filter model leads to engineered single photon loss.  We start with the evolution of a Kerr cat qubit coupled to a series of filter modes.  
\begin{align}
\frac{d\hat{\rho}}{dt}=&-i[-K(\hat{a}^{\dagger 2}-\alpha^2)(\hat{a}^2-\alpha^2),\hat{\rho}(t)]  \nonumber \\
&+ \kappa_1 (1+n_\mathrm{th}) D[\hat{a}]\hat{\rho}(t) + \kappa_1 n_\mathrm{th} D[\hat{a}^\dagger]\hat{\rho}(t)\nonumber \\
&-i[g\hat{f}_1^\dagger \hat{a}e^{i\Delta t}+J\sum_{j=2}^M \hat{f}_{j-1}^\dagger\hat{f}_j + \mathrm{h.c}., \hat{\rho}(t)] \nonumber \\
&+ \kappa_f D[\hat{f}_M]\hat{\rho}(t).
\end{align}
Above we are in the rotating frame of the Kerr cat mode ($\hat{a}$)  and filter modes ($\hat{f}_1, ..., \hat{f}_M$).  In order the terms correspond to the Kerr cat qubit Hamiltonian, the intrinsic single photon loss of the Kerr cat qubit, the intrinsic heating of the Kerr cat qubit, the Hamiltonian coupling to and between the filter modes, and finally the single photon loss of the final filter mode to a cold bath. Also, $\Delta = \omega_f-\omega_a$ represented the frequency detuning between the Kerr cat mode and the filter modes.  To proceed we transform the Kerr cat qubit mode  into the shifted-Fock basis by using the transformation $\hat{a}\simeq\hat{Z}\otimes(\hat{a}' + \alpha)$ in \cref{eq:shifted_fock_annihiliation_operator_sup}. For now we do not include the $\mathcal{O}(e^{-2\alpha^{2}})$ correction in \cref{eq:shifted_fock_annihiliation_operator_sup} (see \cref{sec:filter_low_alpha}). Also, we use an approximate expression for the Kerr cat Hamiltonian $\hat{H}_{\mathrm{KC}}\simeq -4K\alpha^{2}\hat{I}\otimes \hat{a}'^{\dagger}\hat{a}'$. Upon this transformation and approximations, the evolution is described by
\begin{align}
\frac{d\hat{\rho}}{dt}=&-i[-4K\alpha^2 \hat{I}\otimes\hat{a}'^\dagger \hat{a}',\hat{\rho}(t)]  \nonumber \\
&+ \kappa_1 (1+n_\mathrm{th}) D[\hat{Z}\otimes(\hat{a}' + \alpha)]\hat{\rho}(t)\nonumber \\
&+ \kappa_1 n_\mathrm{th} D[\hat{Z}\otimes(\hat{a}'^\dagger + \alpha)]\hat{\rho}(t)\nonumber \\
&-i[g\hat{f}_1^\dagger (\hat{Z}\otimes(\hat{a}' + \alpha))e^{i\Delta t}+J\sum_{j=2}^M \hat{f}_{j-1}^\dagger\hat{f}_j + h.c., \hat{\rho}(t)] \nonumber \\
&+ \kappa_f D[\hat{f}_M]\hat{\rho}(t).
\end{align}
Now we can move into the rotating frame of the Kerr cat shifted-Fock mode ($\hat{a}'$) yielding:
\begin{align}
\frac{d\hat{\rho}}{dt}=&  \kappa_1\alpha^2 (1+2n_\mathrm{th}) D[\hat{Z}\otimes\hat{I}]\hat{\rho}(t)\nonumber \\
&+ \kappa_1 (1+n_\mathrm{th}) D[\hat{Z}\otimes\hat{a}']\hat{\rho}(t) + \kappa_1 n_\mathrm{th} D[\hat{Z}\otimes\hat{a}'^\dagger]\hat{\rho}(t)\nonumber \\
&-i[g\hat{f}_1^\dagger (\hat{Z}\otimes(\hat{a}'e^{i4K\alpha^2 t} + \alpha))e^{i\Delta t}\nonumber \\
&\ \ \ \ \ \ +J\sum_{j=2}^M \hat{f}_{j-1}^\dagger\hat{f}_j + \mathrm{h.c.}, \hat{\rho}(t)] \nonumber \\
&+ \kappa_f D[\hat{f}_M]\hat{\rho}(t).
\end{align}
At this stage we have simplified the dynamics by breaking out terms from the intrinsic loss and heating that lead to  phase flips (see the first two lines). See \cref{sec:leakage_single_photon_loss} for more details on this simplification. We now choose the filter to be detuned by $\Delta = \omega_f - \omega_a=-4K\alpha^2$ (in our numerical simulations though, we observe that $\Delta =-3.6K\alpha^{2}$ works better due to the subleading terms in the Kerr cat Hamiltonian which is ignored here for simplicity). With this choice of detuning the term $\hat{f}_1^\dagger \hat{Z}\otimes\hat{a}'$ which will lead to the desired dissipation is resonant while the undesired term $\alpha\hat{f}_1^\dagger \hat{Z}\otimes\hat{I}'$ that leads to phase-flip errors rotates with a frequency $4K\alpha^2$ ($\omega_a$ in the lab frame).  The difference in rotation frequency between the two terms means that a properly engineered filter can introduce the desired dissipation all the while suppressing the induced phase-flip errors.  In the filter geometry we consider with hopping rate $J$,  the total bandwidth is $4J$.  Hence, the condition $2J < \Delta$ ensures that the undesired phase-flip errors fall outside the filter passband and can be filtered out.  For this section we focus on the desired dissipation so we neglect the off resonant term that leads to phase-flip errors (see \cref{sec:induced_phase_flips}).  Upon adiabatic elimination of all the filter modes the resulting evolution is 
\begin{align}
\frac{d\hat{\rho}}{dt}=&  \kappa_1\alpha^2 (1+2n_\mathrm{th}) D[\hat{Z}\otimes\hat{I}]\hat{\rho}(t)\nonumber \\
&+ \kappa_1 n_\mathrm{th} D[\hat{Z}\otimes\hat{a}'^\dagger]\hat{\rho}(t)\nonumber \\
&+(4g^2/\kappa_f + \kappa_1 (1+n_\mathrm{th})) D[\hat{Z}\otimes\hat{a}']\hat{\rho}(t) . 
\end{align}
This adiabatic elimination of the filter modes requires that $g \ll J$ so that the filter population is low enough to be adiabatically eliminated.  The total cooling rate is $4g^2/\kappa_f + \kappa_1 (1+n_\mathrm{th})$.  
\subsection{Induced Phase-Flip Error Rate}
\label{sec:induced_phase_flips}
Here we show that the induced phase-flip errors are suppressed exponentially with the number of filter modes.  We start with the full dynamics of the Kerr cat qubit and $M$ filter modes in the rotating frame of the filter modes ($\hat{f}_i$) and the gauge mode of the Kerr cat qubit ($\hat{a}')$.
\begin{align}
\frac{d\hat{\rho}}{dt}=& 
-i[g\hat{f}_1^\dagger (\hat{Z}\otimes(\hat{a}'e^{i4K\alpha^2 t} + \alpha))e^{i\Delta t}\nonumber \\
&\ \ \ \ \ \ +J\sum_{j=2}^M \hat{f}_{j-1}^\dagger\hat{f}_j + \mathrm{h.c.}, \hat{\rho}(t)] \nonumber \\
&+ \kappa_f D[\hat{f}_M]\hat{\rho}(t).
\end{align}
We have not included the intrinsic loss mechanisms in this discussion to focus on the induced phase-flip error rate due to the term $g\alpha (\hat{Z}\otimes \hat{I}) \hat{f}_{1}^{\dagger}$.  We also again don't include the exponentially small corrections.  Compared to the situation of the previous section where we adiabatically eliminated the filter modes, here we adiabatically eliminate both the filter modes \textit{and} the shifted-Fock gauge mode so that we can find the dynamics in the qubit sector of the shifted-Fock basis. The adiabatic elimination formalism makes use of a non-Hermitian Hamiltonian describing the evolution of the excited state manifold \cite{ReiterEffective}.  The effective loss operators are related to the non-Hermitian Hamiltonian by $\hat{L}_\mathrm{eff}^{(k)} = \hat{L}_k \hat{H}_\mathrm{NH} \hat{V}_+$  where $\hat{V}_+$ are the Hamiltonian terms that excite the system from the ground to excited manifold and $\hat{L}_k$ are the loss operators of the original problem.  In this specific situation the non-hermitian Hamiltonian describing evolution in the excited states is (the first term can be neglected without changing the scaling)
\begin{align}
\hat{H}_{\mathrm{NH}, \Delta} =&g(\hat{Z}\otimes|1\rangle\langle2|+\hat{Z}\otimes|2\rangle\langle1|)-\frac{i\kappa}{2}\hat{I}\otimes|N\rangle\langle N|\nonumber \\
&+\sum_{j=3}^{N}[J\hat{I}\otimes|j-1\rangle\langle j| + J\hat{I}\otimes|j\rangle\langle j-1|]\nonumber \\
&+\sum_{j=1}^N\Delta[\hat{I}\otimes|j\rangle\langle j|].
\end{align}
We work in the single excitation manifold and use the notation that $|n\rangle$ corresponds to an excitation in the nth filter mode if $n>2$, an excitation in the shifted-Fock mode when $n=1$, and to the ground state when $n=0$.  The other important operators for the adiabatic elimination representing Hamiltonian excitation to the excited state manifold and decay from the excited state manifold are given by 
\begin{align}
V_{+} &=g\alpha \hat{Z}\otimes|2\rangle\langle 0| e^{i\Delta t} \nonumber \\
L_i&=\sqrt{\kappa}\hat{I}\otimes|0\rangle\langle N|
\end{align}
Through adiabatic elimination in the single excitation manifold we find that the dynamics in the qubit sector of the shifted-Fock basis are given by
\begin{align}
\frac{d\hat{\rho}}{dt}=\kappa_{\mathrm{ind}}D[\hat{Z}\otimes\hat{I}]\hat{\rho}(t)
\end{align}
The induced phase-flip rates for 1 and 2 filter modes are given by
\begin{itemize}
\item
1 Filter Mode
\begin{align}
\kappa_{\mathrm{ind}}=\frac{\kappa_f g^2 \alpha^2}{(\Delta-g^2/\Delta)^2 + \kappa_f^2/4}
\end{align}
\item
2 Filter Modes
\begin{align}
\kappa_{\mathrm{ind}}=\frac{\kappa_f g^2 J^2 \alpha^2}{(J^2-\Delta^2+g^2)^2 + (g^2\kappa/\Delta-\kappa\Delta)^2/4}
\end{align}
\end{itemize}
The adiabatic elimination holds in the limit $g \ll J$ and $J\ll \Delta$ which ensure that the population of the filter modes is low enough and that the .  In general the induced phase-flip rate for $M$ filter modes is well described by
\begin{align}
    \kappa_{\mathrm{ind}} = \frac{\kappa_f g^2\alpha^2 J^{2(M-1)}}{\Delta^{2M}}=(4g^2\alpha^{2}/\kappa_{f})\times (J/\Delta)^{2M} \label{eq:kappa induced asymptotic}
\end{align}
again in the limit $g,J \ll \Delta$.
\subsection{Non-adiabatic error suppression due to frequency selective loss}
\label{sec:non adiabatic suppression}
In addition to suppressing leakage and improving the bit-flip rate during idle operation, frequency selective loss can mitigate non-adiabatic gate errors.  Non-adiabatic gate errors refer to predominantly $\hat{Z}$ errors that are induced by rapid gates.  Consider for example the case of the $\hat{Z}$ gate which is implemented using the Hamiltonian $\hat{H}_z = \epsilon (\hat{a}^\dagger + \hat{a})$.  Transforming to the shifted Fock basis and going into the rotating frame of the Kerr cat qubit the Hamiltonian for the gate is $\hat{H}_z = 2i\epsilon\alpha\hat{Z}\otimes\hat{I} + \epsilon \hat{Z}\otimes(\hat{a}'^\dagger e^{i\Delta t} + \hat{a}'e^{-i\Delta t})$.  The first term implements a $Z$ rotation as desired while the second term is undesired and leads to excitations out of the cat state manifold that are accompanied by $Z$ errors. 

When excitations occur due to $\epsilon \hat{Z}\otimes(\hat{a}'^\dagger e^{i\Delta t} + \hat{a}'e^{-i\Delta t})$ the frequency selective single photon loss through the Lindblad term $\kappa_{1, \mathrm{eng}} = D[\hat{Z}\otimes\hat{a}']$ brings the excitations down with another $Z$ rotation.  The net effect of this cycle is then a logical identity in the qubit sector. Hence the non-adiabatic error rates for the $Z$ gate are suppressed.  This is to be contrasted with two-photon dissipation where the cooling comes along with an identity in the qubit sector and the net effect is then a $Z$ rotation leading to $Z$ errors \cite{chamberland2020building}.

Even without this direct canceling of $Z$ errors there is a further benefit due to the frequency selectivity of the loss.  To demonstrate the effect we consider the Hamiltonian $\hat{H} = \epsilon \hat{I}\otimes (\hat{a}'^\dagger e^{i\Delta t} + \hat{a}' e^{-i\Delta t})$.  By changing the operator in the qubit sector of the shifted-Fock basis from a $\hat{Z}$ to an $\hat{I}$ we eliminate the benefit from the previous paragraph and focus on the benefits provided by the frequency selectivity. We can perform the adiabatic elimination of the $M$ filter modes and shifted-Fock mode to find the scaling of this suppression.

\begin{figure}
    \centering
    \includegraphics[width=\columnwidth]{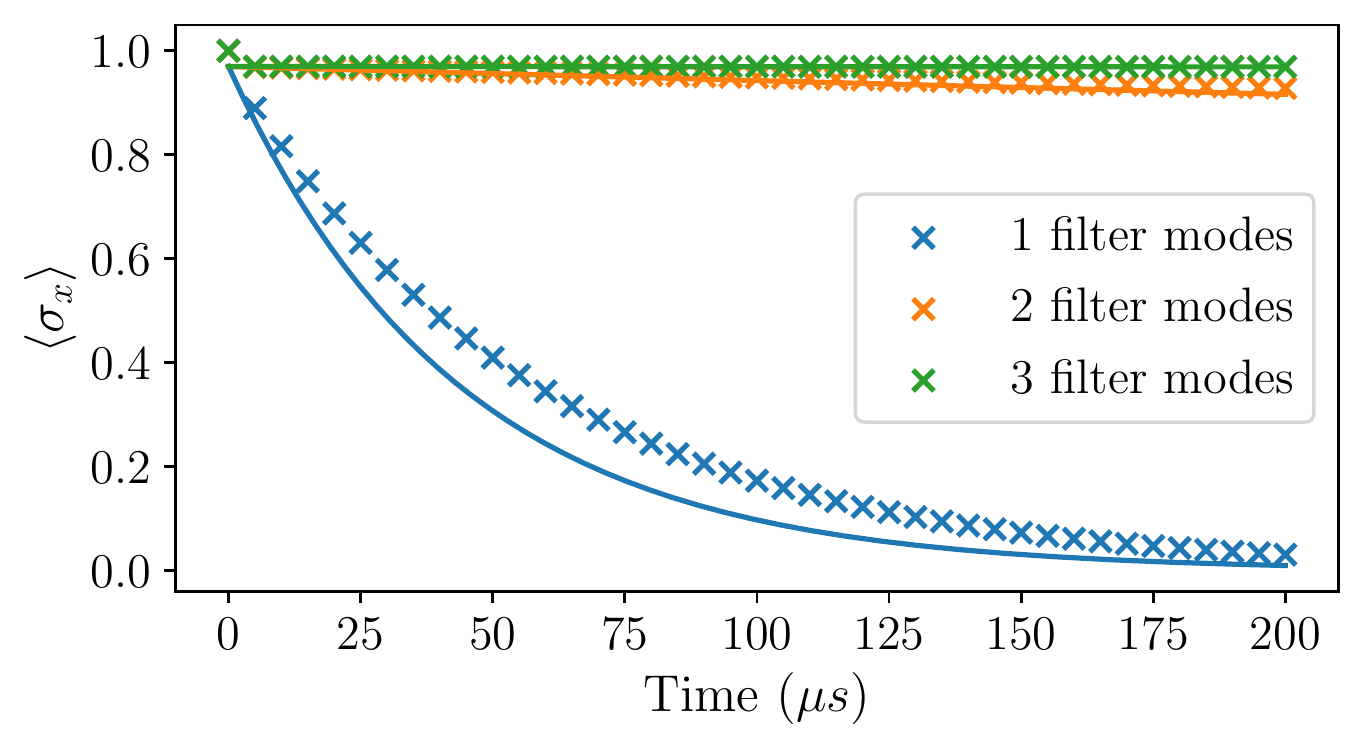}
    \caption{Decay of parity with a Hamiltonian $\hat{H} = \epsilon \hat{I}\otimes( \hat{b}^\dagger + \hat{b})$, initial state $|+\rangle_L$, and varying number of filter modes. The Xs represent numerical simulations and the lines represent the trends expected from the adiabatic elimination.  In this simulation we use a drive strength $\epsilon/2\pi = 30$ MHz.  We explicitly remove the contribution of the decay through the filter modes to isolate the decay due to the drive.  There is an initial jump at short times which is an area for future work.}
    \label{fig:nonadiabatic} 
\end{figure}

To proceed with adiabatic elimination we start with the Lindblad equation describing the colored Kerr cat qubit with the added drive in the rotating frame of the Kerr cat gauge mode ($\hat{a}'$) and filter modes ($\hat{f}_i$)
\begin{align}
\frac{d\hat{\rho}}{dt}&=-i[\epsilon\hat{I}\otimes(\hat{a}'e^{-i\Delta t}+\hat{a}'^\dagger e^{i\Delta t}), \hat{\rho}(t)] \nonumber \\
&-i[g(\hat{f}_1^\dagger (\hat{Z}\otimes(\hat{a}'+\alpha e^{-i4K\alpha^2 t}))+h.c.), \hat{\rho}(t)] \nonumber \\
&-i[J\sum_{j=2}^N (\hat{f}_{j-1}^\dagger\hat{f}_j + h.c.), \hat{\rho}(t)] + \kappa_f D[\hat{f}_N]\hat{\rho}(t) \nonumber
\end{align}
In this case we neglect single photon loss and heating to focus on the effect of the drive induced error.  

To proceed we adiabatically eliminate the gauge mode and the filter modes to find the dynamics in the qubit sector of the shifted-Fock basis.  Compared to the case of the bare Kerr cat qubit there is an additional term which can excite the Kerr cat + filter system to the exicted state manifold given by $\hat{I}\otimes|1\rangle\langle 0| e^{i\Delta t}$. The resulting dissipators for 1 and 2 filter modes are shown below.  The first term in each of the dissipators corresponds to the idling case and the second to the non-adiabatic gate error.
\begin{itemize}
\item
1 Filter Mode
\begin{align}
\frac{d\hat{\rho}}{dt}=D[(&\frac{\sqrt{\kappa_f} g \alpha}{(\Delta-g^2/\Delta) - i\kappa_f/2} +\nonumber \\
&\frac{\sqrt{\kappa_f} g \epsilon}{(g^2-\Delta^2) + i\kappa_f\Delta/2})\hat{Z}\otimes\hat{I}]\hat{\rho}(t)
\end{align}
\item
2 Filter Modes
\begin{align}
\frac{d\hat{\rho}}{dt}=D[&(\frac{\sqrt{\kappa_f} g J \alpha}{(g^2+J^2-\Delta^2) - i(g^2\kappa/\Delta-\kappa\Delta)/2}- \nonumber \\
&\frac{\sqrt{\kappa_f} g J \epsilon}{\Delta(g^2 + J^2-\Delta^2) - i(g^2-\Delta^2)\kappa/2})\hat{Z}\otimes\hat{I}]\hat{\rho}(t)
\end{align}
\end{itemize}
In general the non-adiabatic error due to the drive scales as ${\kappa_f g^2\epsilon^2 J^{2(M-1)}}/{\Delta^{2(M+1)}}$.  Similarly to induced phase-flip errors the non-adiabatic error rate is suppressed exponentially with the number of filters because the drive lies outside of the filter passband.  In \cref{fig:nonadiabatic} we show numerical agreement that by going from 1 to 3 filter modes the non-adiabatic error is suppressed exponentially with the number of filter modes.  Studying both these benefits on more complex gates is an area of future work.  This suppression of off-resonant terms is compatible with two-photon dissipation implemented in a frequency selective manner.

\section{Simulation Details}
\label{sec:simulation_details}
The simulations in this paper were performed using the shifted-Fock basis or a Kerr cat specific basis.  The simulation code was written in Python using the QuTip package.  Unless otherwise specified the simulations use the parameters $\kappa_1/2\pi = 1$ kHz, $n_\mathrm{th} = 0.1$, and $K/2\pi = 10$ MHz.

The shifted-Fock basis was used for Fig. 2 (c) and \cref{fig:nonadiabatic} and is described in more detail Appendix C of \cite{chamberland2020building}.   For the remaining simulations we use a basis specific to the Kerr cat qubits.  The Kerr specific basis is useful because Kerr eigenstates can have non-negligible contributions from highly excited shifted-Fock states which are important to capture to properly find the bit-flip error rate.   The basis states are constructed from the eigenstates of the Kerr cat Hamiltonian $\hat{H} = -K(\hat{a}^{\dagger 2}-\alpha^2)(\hat{a}^{2}-\alpha^2)$ (see also Ref.\ \cite{Xu2021_engineering}). Since the Kerr cat Hamiltonian commutes with parity, the eigenstates can be broken into orthogonal branches with even and odd parity.  Both the shifted-Fock basis and the Kerr-cat basis can be understood as having a sector which represents the qubit information and an gauge sector that captures additional gauge information.  When computing the logical state of the system we trace over the gauge sector and filter sector if the simulation includes filters. 

In simulations we use an gauge dimension of $d_{\max} = 5$ in the Kerr-cat basis.  For increased simulation efficiency the filter modes are not modeled as coupled two level systems. Instead the filter modes are restricted to a single excitation manifold where at most one filter mode is excited.  For example a three mode filter is described by the four basis states $|j_{1}j_{2}j_{3}\rangle$ where $j_{1}+j_{2}+j_{3} \le 1$.  This choice means that the Hilbert space dimension increases by 1 instead of a factor of two with each added filter mode.

\subsection{Bit-flip Rate Fits}
\begin{figure}[htp]
    \centering
    \subfloat{%
      \includegraphics[clip,width=.9\columnwidth]{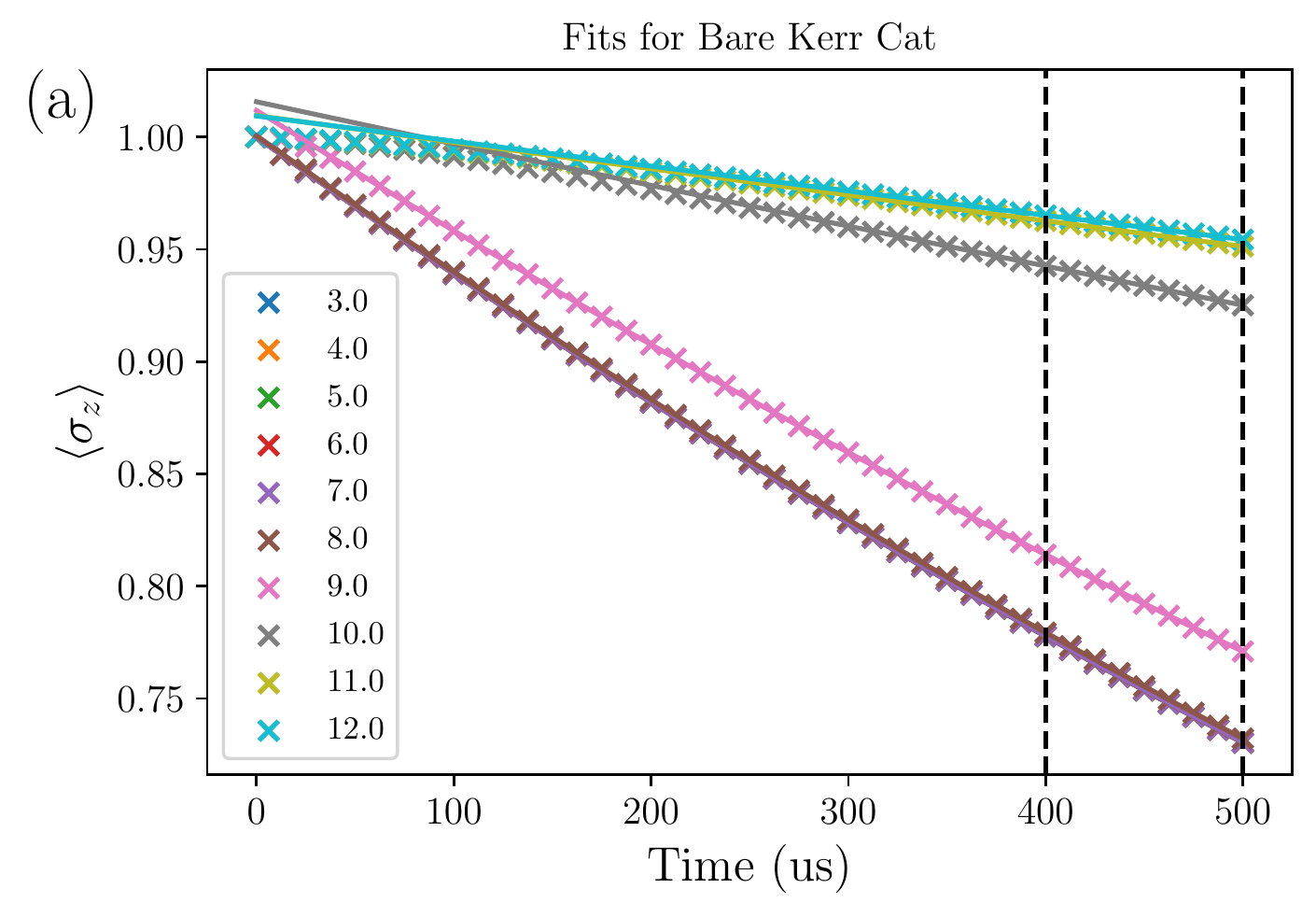}
    }
    
    \subfloat{%
      \includegraphics[clip,width=.9\columnwidth]{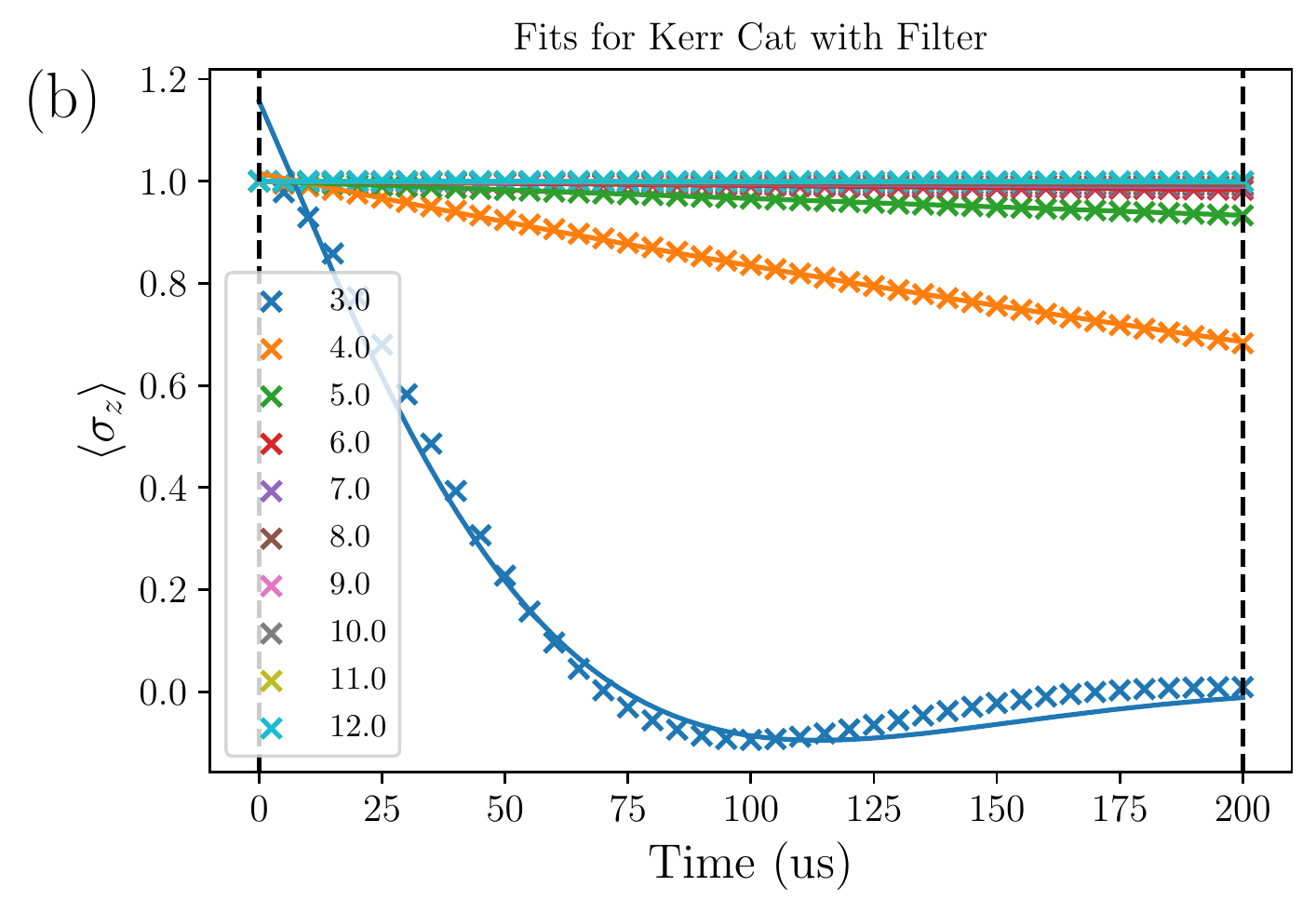}
    }
    
    \caption{ (a) Fits for the bare Kerr cat qubit without any added dissipation.  (b) Fits for the colored Kerr cat qubit with three filter modes.   Both are for the case of a $|0\rangle \simeq |\alpha\rangle$ initial state but the results are very close to symmetric for the case of $|1\rangle\simeq|-\alpha\rangle$. The fit domain in both cases is indicated by the vertical dashed lines.  Not all points are plotted so as to not overcrowd the figure. }
    \label{fig:figKerrBitFlipFits}
\end{figure}

To determine the bit-flip rate we fit the decay of the $\langle \hat{Z}\rangle$ Kerr cat qubit (Here $\hat{Z}$ is the operator on the qubit sector with the gauge sector traced out).  We average over the initial states $|\alpha\rangle$ and $|-\alpha\rangle$ (the symmetry is strong between $|\alpha\rangle$ and $|-\alpha\rangle$ but we average over both to be explicit).  In \cref{fig:figKerrBitFlipFits} we show the fits that yielded the bit-flip rates for Fig. 2 (b).  \cref{fig:figKerrBitFlipFits} (a) shows the fits for a bare Kerr cat qubit (blue curve in Fig 2 (b)).  We fit the decay at long times in order to ensure that we do not undercapture the errors by fitting the short time transient behavior.  A dominant source of the heating from 0 to 2 is the two stage process of heating from 0 to 1 to 2.  Thus in order to accurately predict the totality of this error rate it is essential that the population of the 1 state reach equilibrium.  From Fig. 2 (a), we can see that the 1 state reaches equilibrium on a timescale of roughly $300 \mu s$. Thus we fit the domain $400\ \mu s$ to $500\ \mu s$ which yields converged decay rates for all values of $\alpha^2$ for the bare Kerr cat qubit ($\sim10\%$ different from fitting $300\ \mu s$ to $400\ \mu s$).  

\cref{fig:figKerrBitFlipFits} (b) shows the fits for a Kerr cat qubit coupled to three filter modes (orange curve in Fig 2 (b)).  For $\alpha^2 = 3$ we fit to a decaying sinusoid to get a crude sense of the bit-flip rate.  Further explanation of this small $\alpha^2$ behavior can be found in \cref{sec:filter_low_alpha}.  For the remaining $\alpha^2$ we fit to exponential decays.  Fits are done over the full domain of the simulation.

Fits for the Bare Kerr cat qubit with an added linear drive were done over the domain 240 to 300 $\mu s$.

\subsection{CNOT Simulations}
We simulated the CNOT gate using derivative-based correction for the control pulses as in Ref.\ \cite{Xu2021_engineering}. The Hamiltonian in these simulations includes the Kerr Hamiltonian with drive on the control qubit, the two-mode Hamiltonian that rotates the ground states of the Kerr plus two-photon drive Hamiltonian of the target conditionally on the state of the control, a two-mode control Hamiltonian that accelerates the gate by driving a conditional rotation of the target mode, four derivative-based correction terms, and if colored loss is present also Hamiltonian terms coupling the filter modes and the cat qubit to the filter. The Hamiltonian for the coupling to the filter is given by; 
\begin{align}
    \hat{H} = \hat{H}_{\mathrm{KC}} + \Big{[} g \hat{a}\hat{f}_{1}^{\dagger}e^{i\Delta t} + J\sum_{j=1}^{M-1}\hat{f}_{j}\hat{f}_{j+1}^{\dagger} + \mathrm{h.c.}\Big{]} 
    \label{eq:FilterHamiltonian_sup}
\end{align}
where $\hat{a}$ can be $\hat{a}_c$ or $\hat{a}_t$ which each have their own filter modes.   The Kerr Hamiltonian on the control mode is given by:
\begin{equation}
    \hat{H}_{c} = -K (\hat a^{\dagger 2}_c - \alpha^2)  (\hat{a}_{c}^{2}-\alpha^{2}) .
\end{equation}
If $T$ is the CNOT gate time then let $\Phi(t,T)$ is a function such that $\Phi(0) = 0$ and $\Phi(T) = \pi$. Eventually we will choose $\Phi(t,T)$ to be the integral of a truncated Gaussian. Then
\begin{align}
    \hat{H}_{t} = -K & \left(\hat a_t^{\dagger 2} - \frac{\alpha}{2} (\hat a_c^\dagger + \alpha) + \frac{\alpha}{2} e^{-2i\Phi(t)}(\hat{a}_c^\dagger - \alpha) \right) \nonumber \\
    & \times \left(\hat a_t^2 - \frac{\alpha}{2} (\hat a_c + \alpha) + \frac{\alpha}{2} e^{2i\Phi(t)}(\hat{a}_c - \alpha) \right) . 
\end{align}
The ground state of this Hamiltonian is a cat code in the target mode that rotates if the control mode $\hat{a}_{c}$ is in the logical $|1\rangle_{c}$ state (or approximately $|-\alpha\rangle_{c}$). To perform the gate faster than adiabatically, we use a control Hamiltonian proportional to the derivative of the function $\Phi(T,t)$:
\begin{align}
    \hat{H}_{\mathrm{acc}} = \frac{d \Phi(T,t)}{dt} (\hat a^\dagger_t \hat a_t - \alpha^2)  \frac{1}{4 \alpha}( \hat a_c + \hat a^\dagger_c - 2\alpha) .
\end{align}
The gate given by $\hat{H}_c$, $\hat{H}_t$, and $\hat{H}_{\mathrm{acc}}$ implements a CNOT gate in the cat code manifold. However, if the system has leaked out of the code space to an excited state a wrong unitary is applied. Further, this Hamiltonian drives leakage out of the cat code manifold giving non-adiabatic errors. 

As a concrete example, we break down $\hat{H}_{\mathrm{acc}}$ using the approximate expression for $\hat{a}_{c}$ in the shifted-Fock basis (i.e., $\hat{a}_{c} \simeq \hat{Z}_{c}\otimes (\hat{a}'_{c} + \alpha)$). With this expression, we can see that $\hat{H}_{\mathrm{acc}}$ is approximately given by $\hat{H}_{\mathrm{acc}}^{(1)} + \hat{H}_{\mathrm{acc}}^{(2)}$, where $\hat{H}_{\mathrm{acc}}^{(1)} = -(d\Phi(T,t)/dt) |1\rangle\langle 1|_{c} \otimes (\hat{a}_{t}^{\dagger}\hat{a}_{t}-\alpha^{2}) $ is a desired term that implements the control-qubit-conditional rotation of the target Kerr cat qubit and
\begin{align}
    \hat{H}_{\mathrm{acc}}^{(2)} = \frac{1}{4\alpha} \frac{d\Phi(T,t)}{dt} \hat{Z}_{c}\otimes (\hat{a}'^{\dagger}_{c} + \hat{a}'_{c}) (\hat{a}_{t}^{\dagger}\hat{a}_{t} -\alpha^{2})  
\end{align}
is an undesired term. To accurately analyze this undesired term, we need go to a rotating frame of $\hat{a}_{t}$ and then use the approximate expression $\hat{a}_{t} \simeq \hat{Z}_{t} \otimes (\hat{a}'_{t} + \alpha)$ as was done in Appendix D of Ref.\ \cite{chamberland2020building} as well as in Ref.\ \cite{Xu2021_engineering}. While this procedure affects the qubit sector, the gauge sector is not affected by this frame transformation and is simply given by $( \hat{a}'^{\dagger}_{c}+\hat{a}'_{c} )(\hat{a}'^{\dagger}_{t}\hat{a}'_{t} + \alpha ( \hat{a}'^{\dagger}_{t} + \hat{a}'_{t} ) )$. Thus, we can see that the term $\hat{a}'^{\dagger}_{c}\hat{a}'^{\dagger}_{t}$ induces coherent leakage in both the control and target cat qubits. Also, once the target cat qubit is excited, $\hat{a}'^{\dagger}_{t}\hat{a}'_{t}$ is not trivial and the term $( \hat{a}'^{\dagger}_{c}+\hat{a}'_{c} )\hat{a}'^{\dagger}_{t}\hat{a}'_{t}$ starts to have adverse impacts. Moreover, the excitations in the gauge sectors can be exchanged between the control and target cat qubits through the beam-splitter interaction $\hat{a}'^{\dagger}_{c}\hat{a}'_{t} + \hat{a}'_{c}\hat{a}'^{\dagger}_{t}$. Similarly, the non-linear Hamiltonian on the target Kerr cat qubit $\hat{H}_{t}$ also contains the same beam-splitter interaction which further complicates the dynamics when the system is excited. Due to all such complications associated with leakage, the CNOT Hamiltonians $\hat{H}_{c},\hat{H}_{t},\hat{H}_{\mathrm{acc}}$ implement a wrong unitary whenever the cat qubits are not in their ground state manifolds. Hence, it is important to suppress the leakage.   

As in Ref.\ \cite{Xu2021_engineering} we can reduce coherent leakage using derivative-based corrections. The four first order derivative-based correction Hamiltonian terms are proportional to 
\begin{align}
    H_{\mathrm{DBC}, 0}^{(1)} \propto (\hat a_t^\dagger \hat a_t  - \alpha^2 ) (\hat a_c - \hat a_c^\dagger), 
\end{align}
\begin{align}
    H_{\mathrm{DBC}, 1}^{(1)} \propto \hat a_c + \hat a_c^\dagger, 
\end{align}
\begin{align}
    H_{\mathrm{DBC}, 2}^{(1)} \propto \hat a_c^2 - \hat a_c^{\dagger 2}, 
\end{align}
and
\begin{align}
    H_{\mathrm{DBC}, 3}^{(1)} \propto (e^{2i \Phi(T,t)} - 1)\hat a^{\dagger 2}_t + (e^{- 2i\Phi(T,t)} - 1) \hat a_t^2 .
\end{align}
Each of these terms is engineered to cancel leakage from particular error terms that arise from the evolution under $\hat{H}_c$, $\hat{H}_t$, and $\hat{H}_{\mathrm{acc}}$. Together these derivative-based corrections cancel the first order coherent leakage out of the cat code manifold \cite{Xu2021_engineering}. However, $H_{\mathrm{DBC}, 1}^{(1)}$ and $H_{\mathrm{DBC}, 3}^{(1)}$ have the additional effect of applying a unitary $Z$ rotation to the control qubit. We cancel that rotation at the end by applying $Z(\theta)$ to the control qubit for an angle $\theta$ that is determined by $\frac{d \Phi(T,t)}{dt}$. Note that incoherent leakage (e.g., caused by heating) cannot be suppressed with the pulse shaping and derivative-based correction techniques.

\begin{figure*}
    \centering
    \includegraphics[width=0.9\textwidth]{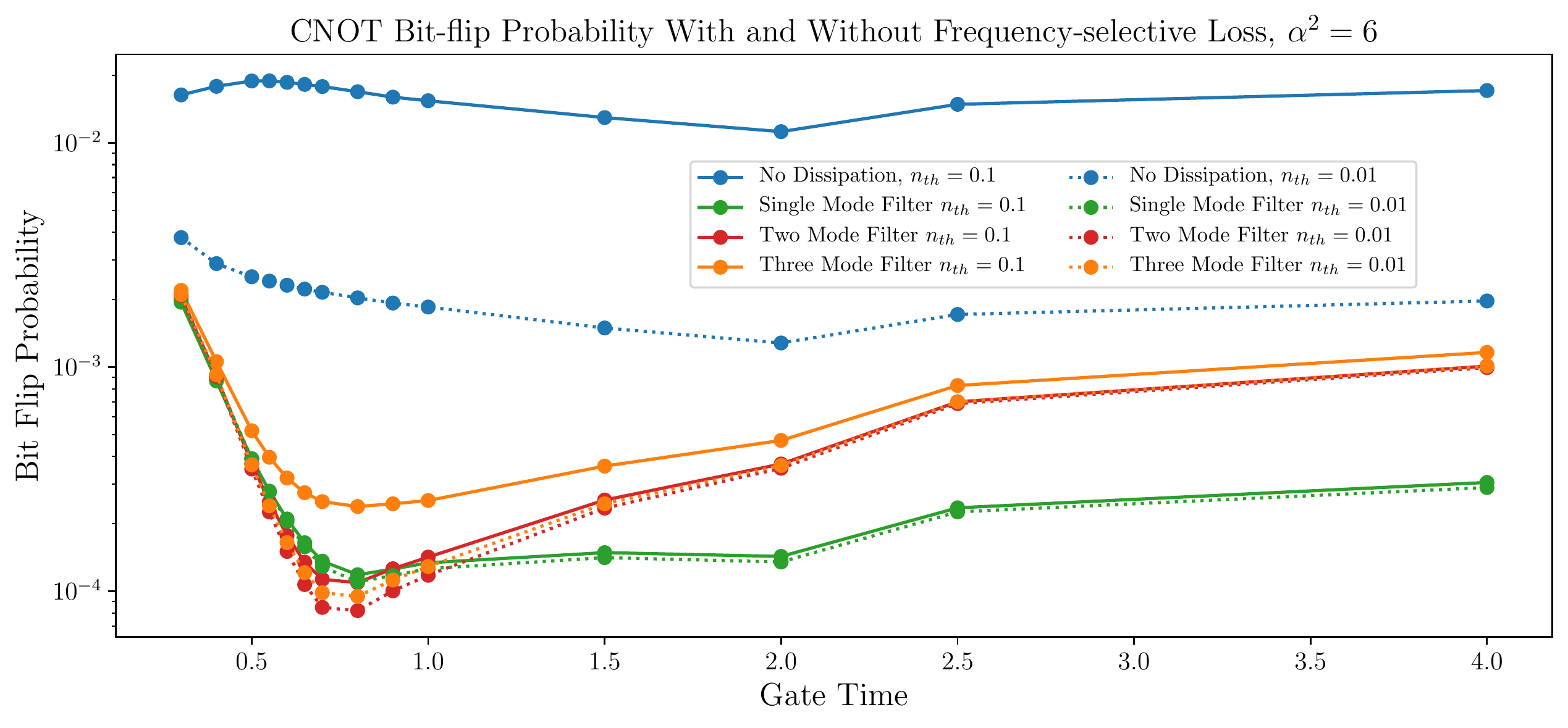}
    \caption{Plot of the bit flip probabilities for the CNOT gate with derivative-based correction for different values of thermal occupation $n_{\mathrm{th}}$. Blue and purple are bare Kerr cat qubits and the other points are colored Kerr cat qubits with between 1 and 3 filter modes. The other parameters are $K = 2\pi \times 10\mathrm{MHz}, \, \kappa_1 = 2\pi \times 1\mathrm{kHZ}$, and $\alpha^2 = 6$.}
    \label{fig:KerrCNOTalpha6}
\end{figure*}

\begin{figure*}
    \centering
    \includegraphics[width=0.9\textwidth]{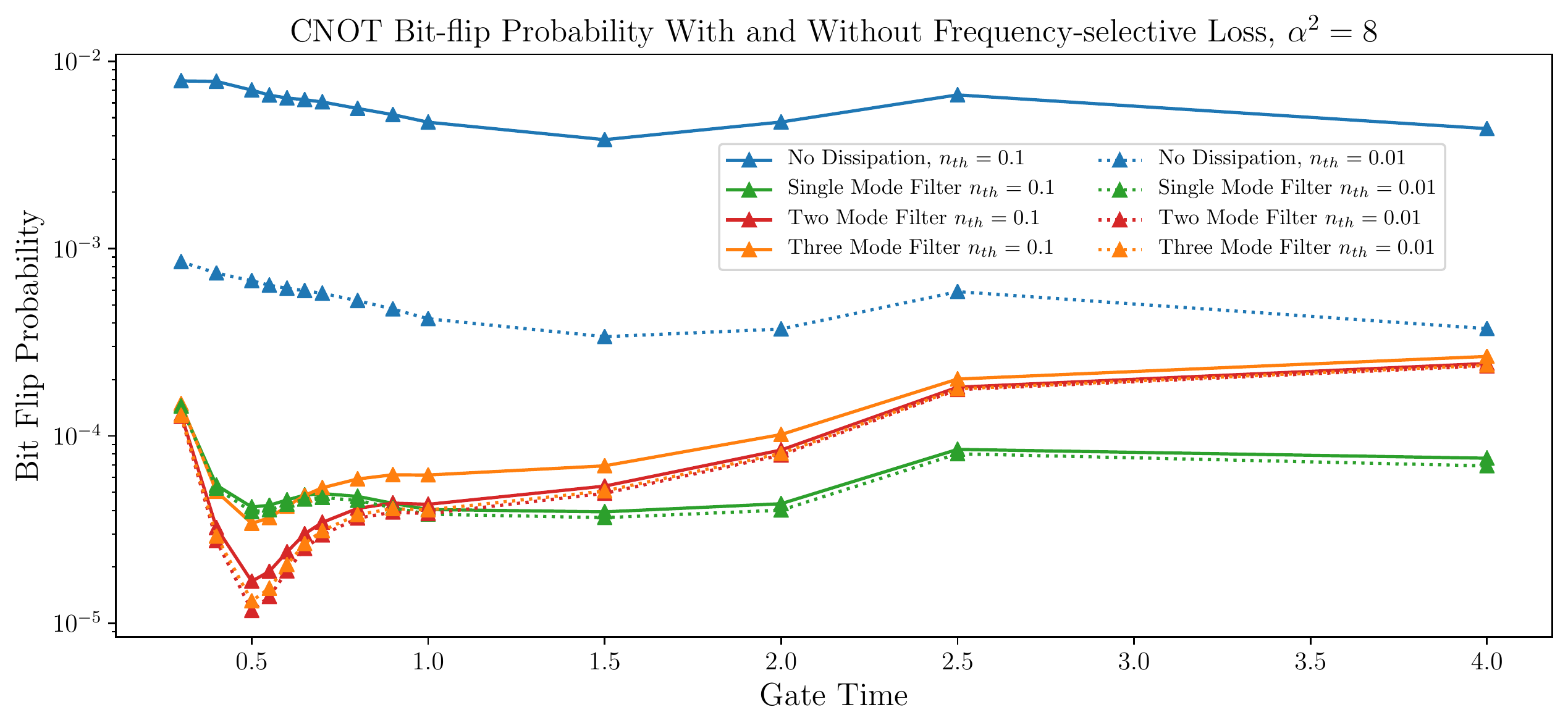}
    \caption{This figure is the same as \cref{fig:KerrCNOTalpha6} but with $\alpha^2 = 8$. This plot shows the bit flip probabilities for the CNOT gate with derivative-based correction for different values of thermal occupation $n_{\mathrm{th}}$. Blue and purple are bare Kerr cat qubits and the other points are colored Kerr cat qubits with between 1 and 3 filter modes. The other parameters are $K = 2\pi \times 10\mathrm{MHz}, \, \kappa_1 = 2\pi \times 1\mathrm{kHZ}$, and $\alpha^2 = 8$.}
    \label{fig:KerrCNOTalpha8}
\end{figure*}

In Fig. 3 we compare the bit-flip performance of the CNOT gate without engineered single-photon loss (i.e., bare Kerr cat qubits) and with single-photon loss filtered by one, two, or three filter modes (i.e., colored Kerr cat qubits). In order to reflect the bit-flip probability of a CNOT gate performed as part of a computation or error correction, we initialized the system in a thermally excited state. Specifically to get the leaked excited state populations for the initial state, the system is prepared in the $|0\rangle_{c} \otimes |0\rangle_{t}$ cat code state (or approximately $|\alpha\rangle_{c} \otimes |\alpha\rangle_{t}$) with the filter in the vacuum state and allowed to evolve under the Kerr Hamiltonian, filter Hamiltonian, loss, and gain. After 300 $\mu s$ the two cat qubits reach a equilibrium population in the gauge sector of their excited states. The excited states of the Kerr cat qubit oscillate between the logical $|0\rangle_{c/t}$ and $|1\rangle_{c/t}$ branches. Therefore, a state with equilibrium excited state population will also have some amount of bit-flip error probability. This initial state, which is primarily in the logical $|0\rangle_{c/t}$ state, is the input to the CNOT simulation. After the gate is complete we trace out the filter and compute the probability of the output state being logical $|0\rangle_{c/t}$ and $|1\rangle_{c/t}$. To account for the initial bit-flip probability we subtract the initial probability of being not in the $|0\rangle_c \otimes |0\rangle_t$ state from the final probability of being not in the $|0\rangle_c \otimes |0\rangle_t$ state. This gives the bit flip probabilities plotted in Fig. 3, \cref{fig:KerrCNOTalpha6,fig:KerrCNOTalpha8} and represents the bit flip probability associated with performing the CNOT gate on the Kerr cat qubit with the equilibrium leakage population. This probability does not accumulate linearly over multiple rounds of CNOT gate because the error is largely due to the initial leakage rather than loss or gain events during the gate. For this reason the bit-flip error probabilities we plot should not be interpreted as parameterizing the stochastic Pauli channel associated with the CNOT gate independent of the initial state. Nevertheless, these error probabilities are what is relevant to computation and error correction where we need to operate on cat states that have evolved long enough to reach an equilibrium excited population in the gauge sector. Our CNOT simulation results underscore that reducing leakage, exactly as colored single-photon dissipation is designed to do, is critical to reducing bit-flip errors in Kerr cat qubits.

\begin{figure*}
    \centering
    \includegraphics[width=0.9\textwidth]{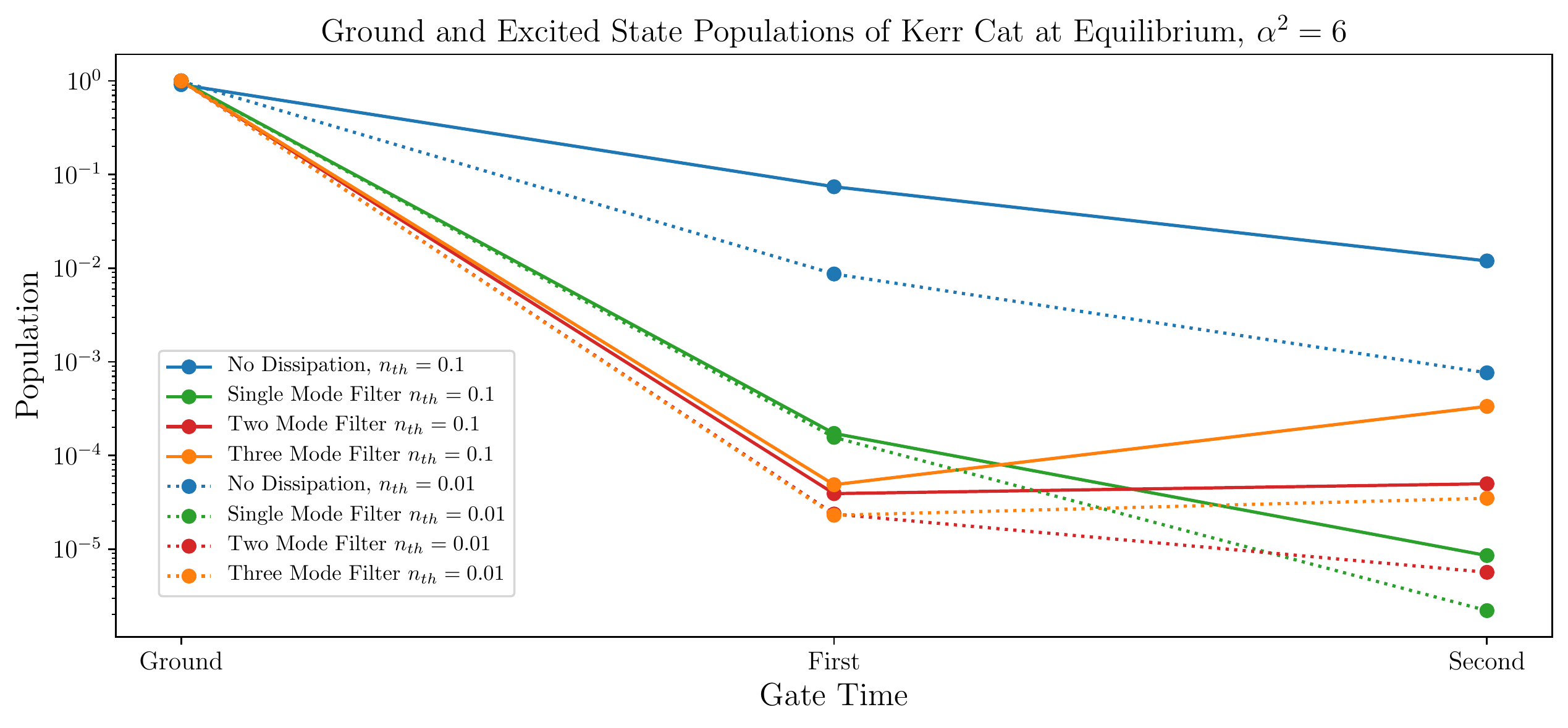}
    \caption{Plot of the populations in the Kerr qubit manifold and the first two excited states after beginning in the code space and idling for $300 \, \mu s$ to reach the equilibrium populations. The populations shown are for the bare Kerr cat qubit as well as the colored Kerr cat qubit with a filter consisting of between one and three modes and for two values of thermal gain, $n_{\mathrm{th}} = 0.1$ and $0.01$. Notice that the second excited state population relative to the first is much smaller for the bare Kerr cat and for the one-mode filter colored Kerr cat than for the two- and three-mode filter colored Kerr cat. Notice also that the first excited state population for $n_{\mathrm{th}} = 0.01$ is much smaller than for $n_{\mathrm{th}} = 0.1$ in the case of the bare Kerr cat but nearly the same for the colored Kerr cat with a one-mode filter. The parameters of the Kerr cat are the same as in the CNOT simulations, $K = 2\pi \times 10\mathrm{MHz}, \, \kappa_1 = 2\pi \times 1\mathrm{kHZ}$, and $\alpha^2 = 6$. These states from idling were used as the input states in the CNOT simulations shown in \cref{fig:KerrCNOTalpha6,fig:KerrCNOTalpha8}.}
    \label{fig:KerrLeakagePopulations}
\end{figure*}

We simulated the CNOT gate for two values of gain $n_{\mathrm{th}} = 0.1$ and $0.01$ and two values of cat state mean photon number $\alpha^2 = 6$ and $8$, for the bare Kerr cat qubit without engineered dissipation and for the colored Kerr cat with a filter of one, two, or three modes. The full set of simulation results from which Fig. 3 is excerpted can be found in \cref{fig:KerrCNOTalpha6,fig:KerrCNOTalpha8}. We worked in the Kerr cat eigenbasis in a rotating frame with gauge dimension of 5. In other words we used the five lowest energy levels around each of the two minima of the Kerr cat potential, hence a total Hilbert space dimension of 10 per cat qubit. The filter was modelled by truncating to the single single excitation manifold so that $M$ filter modes were described by an $M+1$--dimensional Hilbert space. The simulations with three filter modes took about five days each on a single core. The simulations with fewer filter modes or with smaller gauge dimension took hours instead of days. The bit flip probability for $\alpha^2 = 8$ are about an order of magnitude lower than for $\alpha^2 = 6$, reflecting the greater separation between the $|+\rangle$ and $|-\rangle$ coherent states. The dependence on the thermal gain is more complicated. As shown \cref{fig:KerrLeakagePopulations}, as $n_{\mathrm{th}}$ is decreased from $0.1$ to $0.01$, the first excited state population decreases by an order of magnitude for a bare Kerr cat qubit (blue lines). However for colored Kerr cat qubits with one filter mode, the excited state populations decrease only marginally as $n_{\mathrm{th}}$ goes from $0.1$ to $0.01$ (green lines). This is because uncolored single-photon loss may itself introduce some leakage in Kerr cat qubits. See \cref{sec:leakage_single_photon_loss} for a related mechanism. We observe in our numerics that the filters with more modes show greater reductions in the excited state populations as $n_{\mathrm{th}}$ is reduced. If we add up the total leaked population in the four excited states in our simulations, the ratio of the total leakage for $n_{\mathrm{th}} = 0.01$ to the total leakage for $n_{\mathrm{th}}=0.1$ is 0.87 for the one-mode filter, 0.31 for the two-mode filter, and 0.15 for the three-mode filter, almost recovering the ratio of 0.11 for the bare Kerr cat qubit.  Since the leaked population is not a full order of magnitude smaller the bit flip probability in our CNOT simulations for colored Kerr cat qubits when $n_{\mathrm{th}} = 0.01$ is more than 1/10 times the bit flip probability when $n_{\mathrm{th}} = 0.1$. 

Another product of our simplistic filter is that the bit flip probabilities for the three-mode filter are greater than for the one- and two-mode filter. This is because the filter passband is not exactly flat and hence the engineered cooling rate of, e.g., the $|\hat{n}'=2\rangle \rightarrow |\hat{n}'=1\rangle$ transition is not as large as that of the $|\hat{n}'=1\rangle \rightarrow |\hat{n}'=0\rangle$ transition. This leads to greater populations in the higher excited states (i.e., $|\hat{n}'=n\rangle$ with $n\ge 2$), which in turn leads to a larger bit flip probability in our CNOT simulations. For instance, as shown by the red and orange lines in \cref{fig:KerrLeakagePopulations}, the second excited state populations are higher than the first excited state populations for colored Kerr cat qubits with two and three filter modes. However, this issue is specific to the simple filter model we used. Thus, with a more sophisticated, well-optimized filter (yielding a much flatter response within the filter passband) this issue would not appear.

\section{Low $\alpha^2$ Scaling with Filter Modes}
\label{sec:filter_low_alpha}
\begin{figure*}
    \centering
    \includegraphics[width=0.9\textwidth]{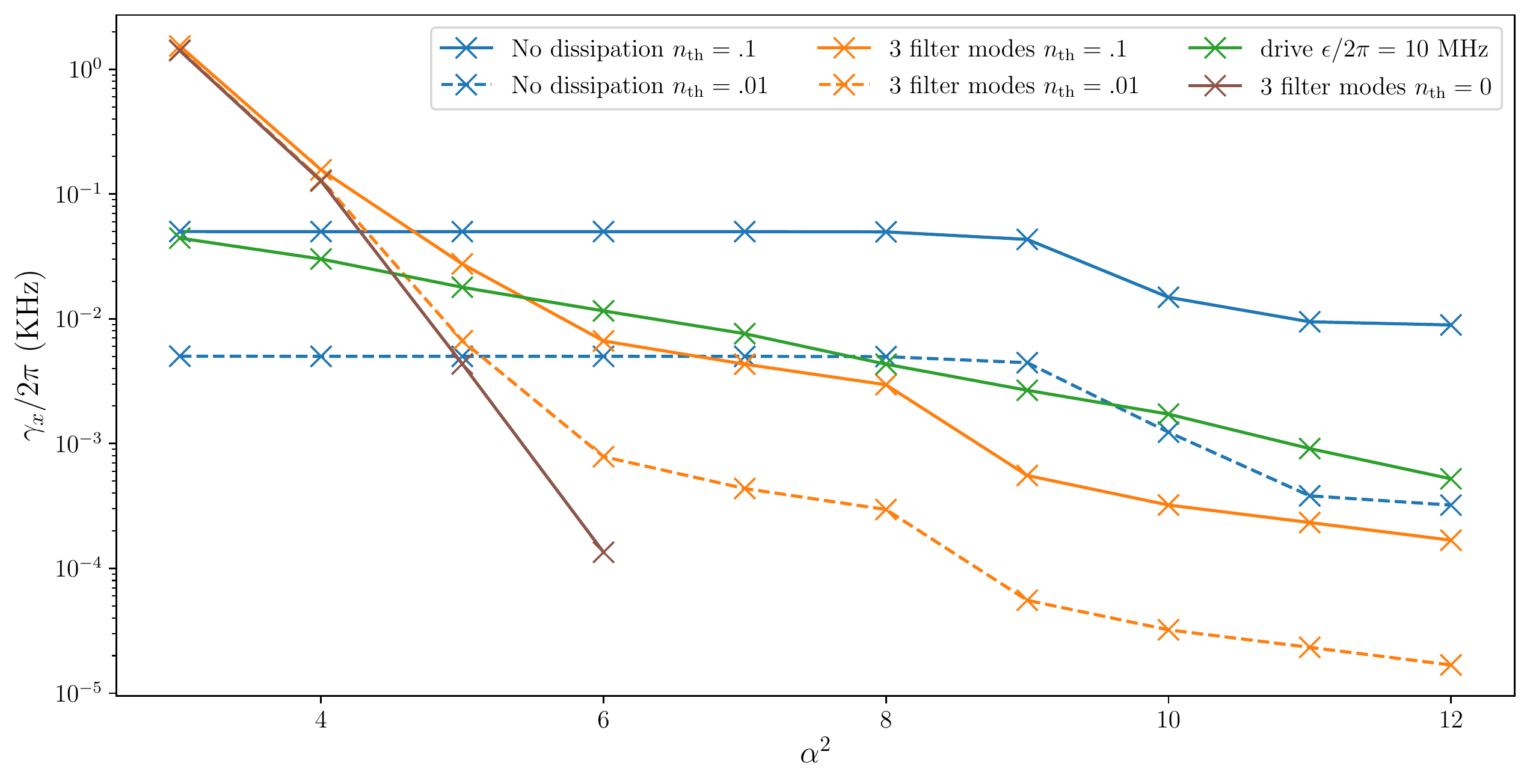}
    \caption{Bit-flip rate vs $\alpha^2$ for different configurations.  In addition to the configurations from Fig. 2 (blue and orange curves), we include the limiting case of a filter with $n_\mathrm{th} = 0$ (brown curve) and a Kerr cat qubit with a linear drive $\epsilon (\hat{a} + \hat{a}^\dagger)$ added (green curve). }
    \label{fig:appendix_bit_flip_numerics}
\end{figure*}

As can be seen in Fig. 2 and  \cref{fig:appendix_bit_flip_numerics} for $\alpha^2 \lesssim 5$ the bit-flip rate of the Kerr cat qubit with added dissipation in the form of a three mode filter is worse than the bare Kerr cat qubit.  The origin of this behavior is the $g\hat{a}^\dagger \hat{f}_1 e^{i\Delta t} + h.c.$ term that couples the first filter mode ($\hat{f}_1)$ to the Kerr cat qubit ($\hat{a}$).  To explain this effect we need to include the exponentially small contributions neglected in \cref{sec:filter_elimination}.  In the shifted-Fock basis we can express the lowering operator as $\hat{a} = \hat{Z}\otimes(\hat{a}'+\alpha) + i\hat{Y}\otimes{\hat{a}_y}$ where $\hat{a}_y$ scales as $\mathcal{O}(e^{-2\alpha^2})$ for the explicit expression of $\hat{a}_{y}$, see \cref{eq:shifted_fock_annihiliation_operator_sup}.  To understand the behavior of the bit-flip rate for small $\alpha^2$ the $\hat{Z}\otimes\hat{a}'$ term is not relevant because it only produces the desired dissipation of the Kerr cat qubit mode.  The relevant terms are $\hat{Z}\otimes{\alpha}$ and $\hat{Y}\otimes\hat{a}_y$. From the latter term we focus on $\xi\hat{Y}\otimes  |\hat{n}'=0\rangle\langle \hat{n}'=0| \simeq \alpha e^{-2\alpha^2}\hat{Y}\otimes|\hat{n}'=0\rangle\langle\hat{n}'=0|$ because the Kerr oscillator is predominantly in its ground state manifold.  The effective Hamiltonian for the logical information in the limit of low filter population is then
\begin{align}
    \hat{H}_{\mathrm{eff}} \approx -\frac{2 g^2 \xi \alpha}{\Delta} \hat{X} \otimes |\hat{n}'=0\rangle\langle \hat{n}' =0|
\end{align}
and the full dynamics in the qubit sector of the shifted-Fock basis are well described by the Lindblad equation
\begin{align}
    \frac{d \hat{\rho}(t)}{dt}\approx-i[\hat{H}_{\mathrm{eff}}, \hat{\rho}(t)] + (\kappa_1 \alpha^2 + \kappa_{\mathrm{ind}})D[\hat{Z}]\hat{\rho}(t).
\end{align}
See \cref{eq:kappa induced asymptotic} for the definition of $\kappa_{\mathrm{ind}}$. With three filter modes $\kappa_{\mathrm{ind}} \ll \kappa_1 \alpha^2$ as was shown in Fig. 2 (c) so we focus on $\kappa_1 \alpha^2$.  Intuitively when $\kappa_1=0$ the Kerr cat qubit undergoes coherent $\hat{X}$ rotations.  When $\kappa_1 \alpha^2$ is non-zero the rotations become incoherent leading to uncorrectable bit-flip errors.  When the phase-flip error rate is much larger than the Hamiltonian rate the bit-flip rate is suppressed by the stochastic $Z$ rotations.  For $\alpha^2 \sim 3$ the period of the $X$ rotations is $\sim 150 \mu s$ while $\kappa_1 \alpha^2 \sim 1/50 \mu s$.  Hence $\alpha^2 \sim 3$ ends up being a heavily damped case. 

We have run numerics to check that this loss mechanism does not limit the single photon frequency selective loss for lower values of $n_\mathrm{th}$.  As can be seen in \cref{fig:appendix_bit_flip_numerics} even with $n_\mathrm{th} = 0.01$ the bit-flip rate is still suppressed by roughly an order of magnitude at $\alpha^2 \sim 6$.  In the brown curve we show the limit imposed on the bit-flip time due to the presence of the filter as a function of $\alpha^2$ by finding the bit-flip rate with $n_\mathrm{th} = 0$.  At $\alpha^2 \sim 7$ there are a few orders of magnitude between the bit-flip rate due to the thermal heating with $n_\mathrm{th} = 0.01$ and the effective Hamiltonian.

This loss mechanism is also mitigated by the dynamical decoupling discussed in \cref{sec:dynamical_decoupling}.

\section{Leakage Inherent to Single Photon Loss}
\label{sec:leakage_single_photon_loss}
Here we explain how uncolored single photon loss stabilizes some leakage in a Kerr oscillator.  Uncolored single photon loss is described by the evolution $\frac{d\hat{\rho}}{dt}=\kappa_1 D[\hat{a}]\hat{\rho}$.  Converting to the shifted-Fock basis the evolution is described by $\frac{d\hat{\rho}}{dt}=\kappa_1D[\hat{Z}\otimes(\hat{a}'+\alpha)]\hat{\rho}$.  We can now expand out the Lindlad equation to reexpress this in an alternative form
\begin{align}
    \frac{d\hat{\rho}}{dt}=&\kappa_1D[\hat{Z}\otimes(\hat{a}'+\alpha)]\hat{\rho} \nonumber \\
    =& \kappa_1(\hat{Z}\otimes(\hat{a}'+\alpha)^\dagger\hat{\rho} (\hat{Z}\otimes(\hat{a}'+\alpha)\nonumber \\
    &-\frac{\kappa_1}{2}\left\{(\hat{Z}\otimes(\hat{a}'+\alpha)^\dagger (\hat{Z}\otimes(\hat{a}'+\alpha), \hat{\rho}\right\}\nonumber \\
    =&\kappa_1 D[\hat{Z}\otimes\hat{a}']\hat{\rho}+\kappa_1 \alpha^2 D[\hat{Z}\otimes\hat{I}]\hat{\rho}\nonumber \\
    &+\kappa_1\alpha\left(\hat{Z}\otimes\hat{a}'^\dagger\hat{\rho}\hat{Z}\otimes\hat{I} +\hat{\rho}\hat{Z}\otimes\hat{I}\hat{\rho}\hat{Z}\otimes\hat{a}'\right)\nonumber \\
    &-\frac{\kappa_1\alpha}{2}\left(\hat{I}\otimes(\hat{a}'+\hat{a}'^\dagger)\hat{\rho}+\hat{\rho}\hat{I}\otimes(\hat{a}'+\hat{a}'^\dagger)\right)\nonumber \\
    =&\kappa_1 D[\hat{Z}\otimes\hat{a}']\hat{\rho}+\kappa_1 \alpha^2 D[\hat{Z}\otimes\hat{I}]\hat{\rho}\nonumber \\
    &-i\left[\frac{i\kappa_1\alpha}{2}(\hat{a}'^\dagger-\hat{a}), \hat{\rho}\right]
\end{align}
To go from the third to fourth step we assume that the the qubit sector in the shifted-Fock basis is in a maximally mixed state so that we can move the $\hat{Z}$ through $\hat{\rho}$.  In reality for the Kerr cat qubit there is also a detuning of the drive by $\Delta\sim 4K\alpha^2$ in the shifted-Fock basis so that the total evolution is given by
\begin{align}
    \frac{d\hat{\rho}}{dt}=&\kappa_1 D[\hat{Z}\otimes\hat{a}']\hat{\rho}+\kappa_1 \alpha^2 D[\hat{Z}\otimes\hat{I}]\hat{\rho}\nonumber \\
    &-i\left[\frac{i\kappa_1\alpha}{2}(\hat{a}'^\dagger-\hat{a}') +\Delta \hat{a}'^\dagger \hat{a}', \hat{\rho}\right]
\end{align}
Thus we see that roughly the presence of uncolored single-photon loss can lead to the stabilization of a coherent state of amplitude ${\kappa_1 \alpha}/{2\Delta}$ in the gauge sector of the shifted-Fock basis due to the last term which is a detuned drive.  The stabilization of a coherent state in the gauge sector of the shifted-Fock basis means that there is an additional baseline amount of leakage.  Nonetheless as we show in Fig. 2 when using three filter modes as opposed to the uncolored loss there is still significant improvement in advancing from $n_\mathrm{th} = 0.1$ to $n_\mathrm{th} = 0.01$ which one could fear would be limited by this.

\section{Hamiltonian Engineering}
\label{sec:dynamical_decoupling}

In this section we explain how adding Z rotations (parity oscillations) to Kerr cat qubits can suppress bit-flip errors.  

As was discussed bit-flip errors in Kerr cat qubits can be dominantly attributed to leakage to excited eigenstates of the Kerr oscillator which have strong interwell couplings.  These couplings are strong enough that an excitation has a large probability of leading to a bit-flip error.  The coupling terms take the form 
\begin{align}
    \hat{H}_\mathrm{couple}\simeq\sum_n \chi_n \hat{X}\otimes|n\rangle\langle n|.
\end{align}
In the qubit sector the $\hat{X}$ indicates a coupling between the wells of the Kerr cat qubit and in the gauge sector $|n\rangle\langle n|$ indicates the coupling is between equally excited states in both wells.

The dynamics of the Kerr cat qubit that lead to bit-flip errors are a combination of incoherent heating and unitary evolution under the interwell coupling Hamiltonian.  When the Kerr cat qubit is excited to level $|n\rangle$ in the gauge sector, the qubit sector evolution is described by the unitary $\hat{U}(t) = e^{i\chi_n \hat{X}t} = \hat{I}\cos{(\chi_n t)} + i\hat{X}\sin{(\chi t)}$.  In a time $t = \pi/2\chi$ a bit-flip will have occurred.  If $\kappa_n \ll \chi_n$ many rotations will occur and the logical $Z$ information will be scrambled.

One way to view the addition of $Z$ rotations is somewhat analogously to the addition of $\pi$ pulses to suppress dephasing.  The $Z$ rotations result in a constant change of the direction of the $X$ rotations so that they interfere on themselves.  In essence we are performing dynamical decoupling in the qubit sector of the shifted Fock basis where the X rotation will continually refocus the Z rotation to lower the chances of a full rotation occurring.  An important distinction between these two situations is that in the Kerr cat case the $Z$ rotations suppress a \textit{Hamiltonian} coupling activated by jump heating. This is to be contrasted with the addition of $\pi$ pulses to directly suppress jump dephasing.

Alternatively we can directly investigate the qubit-sector dynamics of the Kerr cat qubit.  With the drive added, the rotation in the qubit sector will be described by $\hat{U}(t) = \hat{I}\cos{(\gamma t)} + i(\frac{\epsilon}{\gamma}\hat{Z} +\frac{\chi}{\gamma} \hat{X})\sin{(\gamma t)}$ where $\gamma = \sqrt{\epsilon^2+\chi^2}$.  In this form we can see that large $\epsilon$ has the effect of minimizing the scale of the induced rotation.  With the interwell coupling off resonance the bit-flip error probability due to an excitation to level $n$ is upper bounded by $(\chi_n/\gamma)^2$.

This suppression of the bit-flip rate does not need to be implemented with a continuous drive.  More general pulses sequences of $Z$ rotations would also mitigate leakage induced bit-flips by decoupling the higher levels.  For example with a sequence of delta function $Z$ rotations with spacing of $1/\Delta$ the bit-flip error probability is upper bounded by $\sin^2{(\chi/\Delta)} \approx (\chi/\Delta)^2$ where the approximation holds in the limit $\chi \ll \Delta$. 

In \cref{fig:appendix_bit_flip_numerics} the green curve shows the bit-flip time of the Kerr cat qubit when a continuous drive of strength $\epsilon = 2\pi \times 10$ MHz is added to the Kerr Hamiltonian.  At the experimentally relevant values of $\alpha^2 \gtrsim 6$ there is over an order of magnitude improvement over the bare Kerr cat qubit.

While the dynamical decoupling technique can suppress the bit-flip errors in bare Kerr cat qubits, leakage cannot be suppressed by this technique. In principle, however, the dynamical decoupling can be combined with the frequency selective single photon loss to suppress leakage and yield further improved bit-flip error rate. If one uses a continuous drive on the Kerr cat qubit to improve the bit-flip time the amount of tunneling is suppressed but the oscillation rate (now $\gamma\approx\epsilon$) is enhanced.  In this circumstance the single photon loss will add little benefit because $\epsilon \sim \kappa_{1, \mathrm{eng}}$.  Thus it is better to add $Z$ rotations as echo pulses.  In the intermittent time between the pulses the Kerr cat will benefit with the shorter lifetime of excitations.  Furthermore the frequency selective loss will mitigate the non-adiabatic gate error from the physical implementation of a $Z$ rotation. We leave such combination of dynamical decoupling and frequency-selective single-photon loss as a future work.

\section{Calculations relating to Bit-flip rate of Kerr Cat Qubits}
\label{section:Kerr-cat-bit-flip}

\subsection{Kerr Cat Eigenbasis}
\label{sec:kerr_cat_eigenbasis}
Here we compute the leading order perturbative eigenstates of the Kerr oscillator.  The Kerr cat Hamiltonian in the shifted Fock basis is approximately
\begin{align}
\hat{H}_{\mathrm{KC}} &= -4K\alpha^2 \hat{I} \otimes  \hat{a}'^\dagger \hat{a}' - K\hat{I}\otimes \hat{a}'^{\dagger 2}\hat{a}'^2
\nonumber\\
& \!\!\!\! - 2K\alpha\hat{I}\otimes (\hat{a}'^{\dagger 2}\hat{a}' +  \hat{a}'^\dagger) \hat{a}'^2 + \mathcal{O}( e^{-2\alpha^{2}} ).
\label{eq:KC_Hamiltonian}
\end{align} 
Treating $\hat{I}\otimes- 2K\alpha (\hat{a}'^{\dagger 2}\hat{a}' + \hat{a}'^2 \hat{a}'^\dagger)$ as a perturbation to the Hamiltonian $\hat{H}_0 = \hat{I} \otimes [-4K\alpha^2 \hat{a}'^\dagger \hat{a}' - K\hat{a}'^{\dagger 2}\hat{a}'^2]$ we can compute the gauge sector of the eigenstates to leading order in perturbation theory: 

\begin{align}
|\hat{n}'' = n\rangle &= \frac{\alpha (n-1)\sqrt{n}}{2\alpha^2 + (n-1)}|\hat{n}' = n-1\rangle  \nonumber\\
&+ |\hat{n}' = n\rangle - \frac{\alpha n\sqrt{n+1}}{2\alpha^2 + n}|\hat{n}' = n+1\rangle + O((1/\alpha)^2), 
\end{align}
where $n\ge 1$. Here $|\hat{n}' = n\rangle$ represents the gauge sector in the shifted-Fock basis. The ground state in the new basis is the same as the original basis i.e. $|\hat{n}'' = 0\rangle = |\hat{n}' = 0\rangle$.  

Considering the example of the first Kerr cat eigenstate we see that it is predominantly described by the first shifted-Fock excited state.  Nonetheless to more accurately predict the tunneling rate of the first Kerr eigenstates it is important to include the contributions of the higher shifted-Fock excited states.

\subsection{Kerr Basis Tunneling rate}
The rate of tunneling $\chi_n'$ through the n-th excited level, or the half energy splitting between the n-th pair of excited states of the Kerr-cat quit, can be roughly estimated by approximating the Hamiltonian \cref{eq:KC_Hamiltonian} as an oscillator with double-well potential, which is a well studied model whose energy levels can be calculated by, e.g. WKB approximation. However, $\chi_n'$ obtained in this way is not very accurate since the high-order terms of the momentum quadrature is neglected in the double-well approximation. Here to more accurately describe $\chi_n'$, we use the following empirical expression
\begin{equation}
    \chi_n' \approx K e^{-1.6 (\alpha^2 - 4 n)},
\label{eq:tunnelling_rates_expression}
\end{equation}
which agrees well with the numerical values for the states below the potential barrier $n < \alpha^2/4$ (or in the regime $\chi_n'/K < 1$). 

\begin{figure}
    \centering
    \includegraphics[width=0.9\columnwidth]{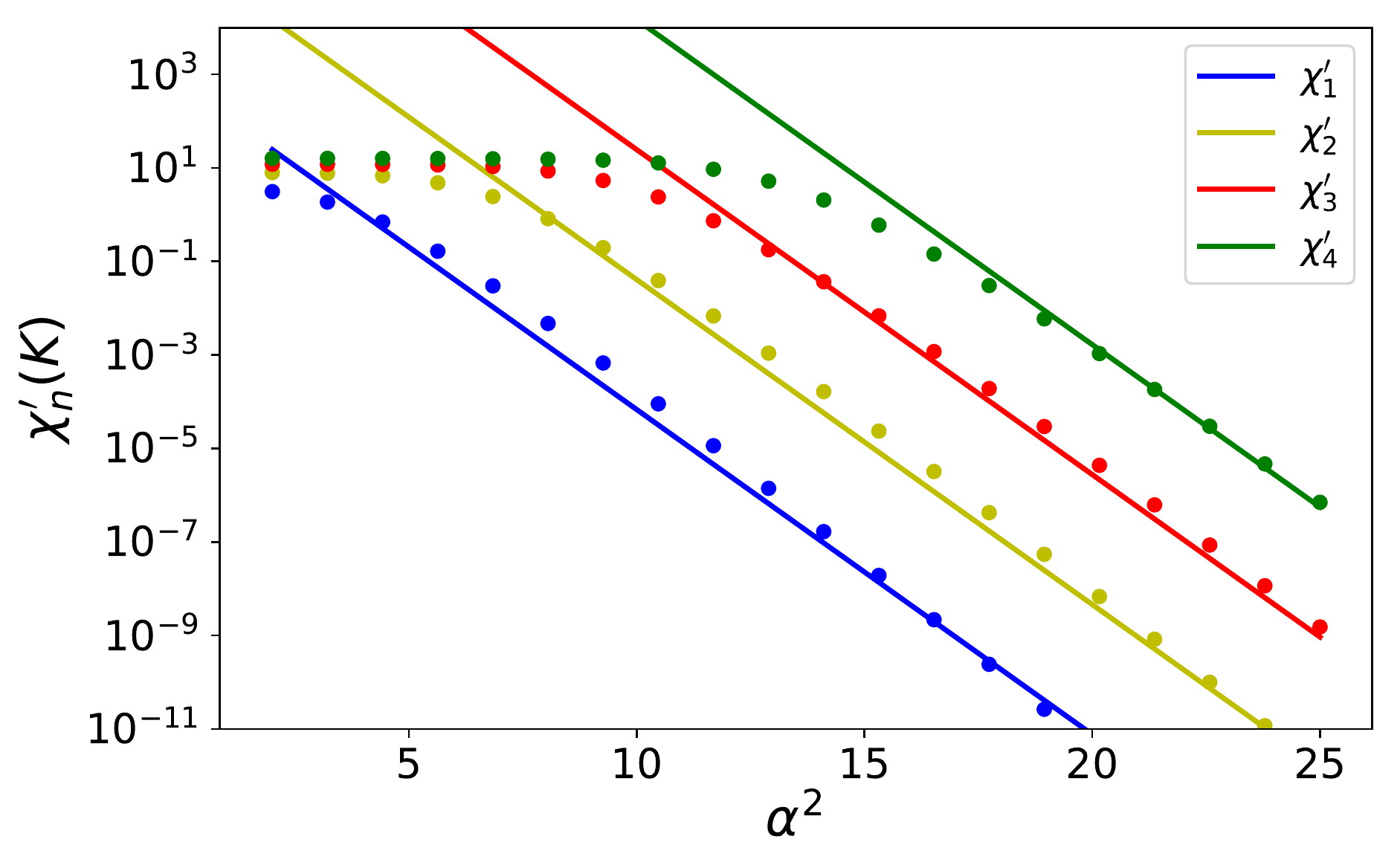}
    \caption{Plot of the tunneling rates (or the interwell coupling) of the first 4 Kerr-cat excited state ($\chi_n^{\prime}$) in units of $K$. The dots indicate numerically extracted tunneling rates while the solid lines indicates the empirical expression by \cref{eq:tunnelling_rates_expression}. Good agreement is shown between the numerical values and the empirical expression \cref{eq:tunnelling_rates_expression} in the regime $\chi_n^{\prime}/K < 1$. }
    \label{fig:tunneling_rates_Kerr_Basis}
\end{figure}

The numerically extracted tunneling rates $\chi_n'$ for the first and second excited states, in comparison with the expression \cref{eq:tunnelling_rates_expression}, can be seen in \cref{fig:tunneling_rates_Kerr_Basis}.

\subsection{Heating/Cooling Rate}
Here we calculate the heating rate to and the decay rate from the excited states of the Kerr-cat qubit. The heating that we consider dominantly comes from the coupling to the thermal bath, which is described by the dissipator $n_{\mathrm{th}}\kappa_1 \mathcal{D}[\hat{a}^{\dagger}]$, where $n_{\mathrm{th}}$ is the thermal occupation of the bath. 

To calculate the heating rate to the n-th pair of excited states we can decompose $\hat{a}^{\dagger}$ in the Kerr-cat eigenbasis:
\begin{equation}
    \hat{a}^{\dagger} = \hat{Z}\otimes [|\hat{n}''=1\rangle \langle \hat{n}''=0| + \sum_{n=2} \lambda_{0,n}|\hat{n}''=n\rangle \langle \hat{n}''=0|],
\end{equation}
where the matrix elements $\lambda_{0,n}$ can be perturbatively calculated as:
\begin{equation}
\lambda_{0,n} = \sqrt{n!} \left(\frac{1}{2 \alpha}\right)^{n-1} + O((\frac{1}{\alpha})^{n}).
\end{equation}
Here for simplicity we only show the matrix elements of $\hat{a}^{\dagger}$ that connect the ground states to excited states.

For small $n_{\mathrm{th}}$, the heating to the n-th pair of excited states dominantly comes from the direct heating from the ground states, and therefore the rate of heating to excited states is approximately given by:
\begin{equation}
    \kappa_{0 \rightarrow n} \simeq  n_{\mathrm{th}}\kappa_1 |\lambda_{0,n}|^2.
\end{equation}

Assuming an ideal engineered dissipation rate $\kappa_{1, \mathrm{eng}}$, i.e. the engineered dissipation is described by the dissipator $\kappa_{1, \mathrm{eng}} \mathcal{D}[\hat{Z}\otimes \hat{a}']$, the decay of excited states is dominated by the decay between adjacent levels, therefore the excited states' decay rate is approximately given by:
\begin{equation}
    \kappa_{n \rightarrow n - 1} \simeq  n \kappa,
\end{equation}
where $\kappa = (\kappa_1 + \kappa_{1, \mathrm{eng}})$.

\subsection{Bit-flip error rate}
At this point we have the heating rate $\kappa_{0\rightarrow n}$, the tunneling rate $\chi_n$, and the lifetime $1/\kappa_{n\rightarrow n-1}$ of each pair of the excited states. We can then estimate the total bit-flip rate by summing up the population that tunnels through the potential barrier from each energy level per unit time:
\begin{equation}
    \gamma_X = \sum_{n = 1} \gamma_X^{(n)}
    \label{eq:bit-flip-analytics}
\end{equation}
where 
\begin{equation}
    \gamma_X^{(n)} = \left\{\begin{array}{ll}
\kappa_{\uparrow, n}/2, & \text { if } \chi_n'/\kappa_{\downarrow,n}>\pi/4 \\
\kappa_{\uparrow, n} \sin^2 \chi_n'/\kappa_{\downarrow,n}, & \text { if } \chi_n'/\kappa_{\downarrow,n}<\pi/4 
\end{array}\right.
\end{equation}

\begin{figure}[h!]
    \centering
    \includegraphics[width=0.45\textwidth]{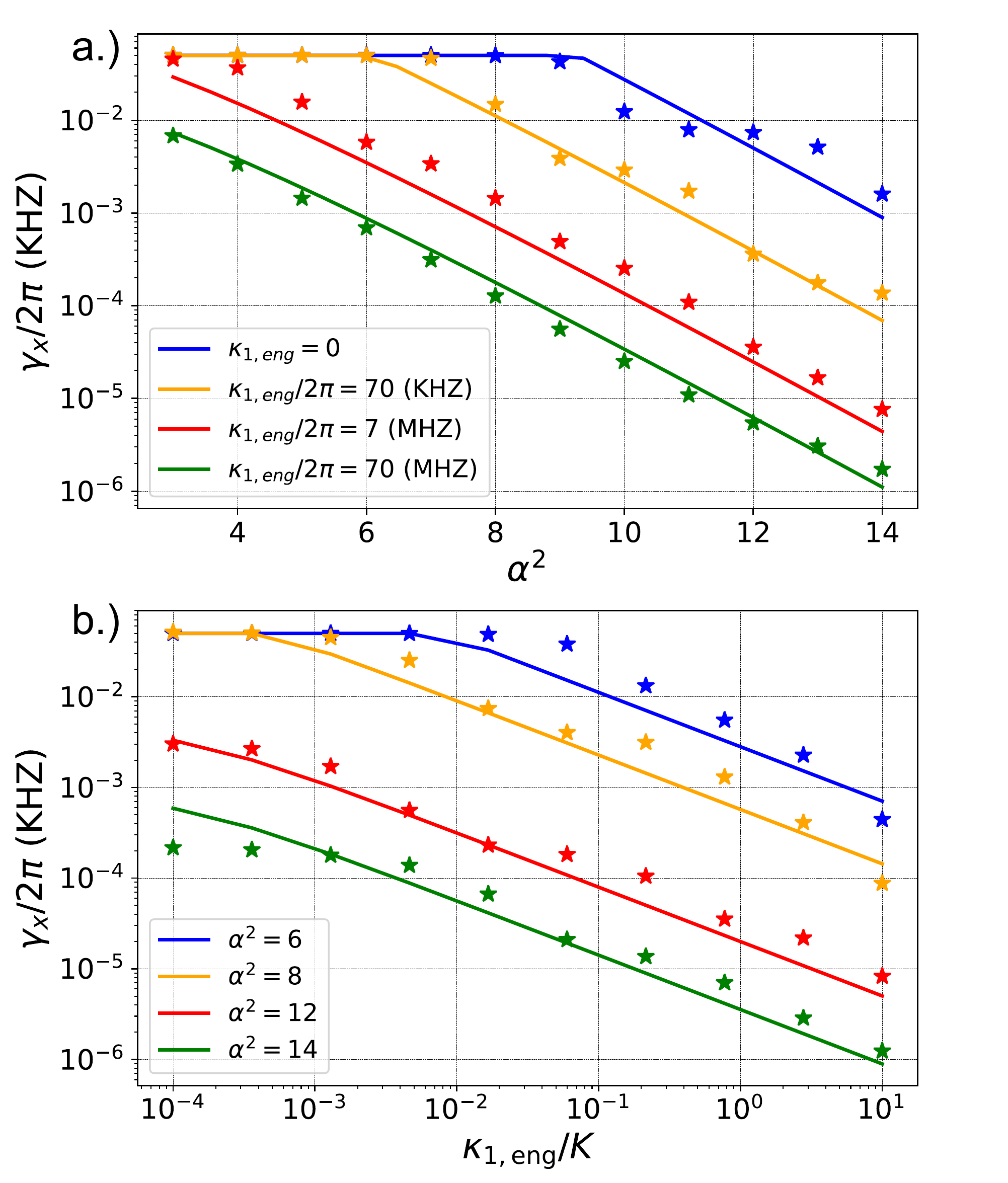}
    \caption{(a). Bit-flip rates as a function of $\alpha$ for different sets of $\kappa_{1,\mathrm{eng}}$. (b). Bit-flip rates as a function of $\kappa_{1,\mathrm{eng}}$ for different sets of $\alpha$. Stars are numerically extracted values while solid lines are theoretical values given by \cref{eq:overall_bit_flip_rate}. Good agreement is shown between \cref{eq:overall_bit_flip_rate} and the nuemrical results.}
    \label{fig:bit_flip_analytics}
\end{figure}

The contribution to the total bit flips dominantly comes from the "critical" excitation level $n_c$ when $\chi_{n_c}^{\prime}$ is comparable to $\kappa_{n_{c} \rightarrow n_{c}-1}$. Therefore, we can give an approximation estimation for $\gamma_X$ by only taking into account the critical level $n_c$:
\begin{equation}
    \gamma_X \approx \gamma_X^{(n_c)} = \kappa_1 n_{\mathrm{th}}|\lambda_{0, n_{c}}|^2 /2 = \kappa_1 n_{\mathrm{th}} (n_c!) (\frac{1}{4\alpha^2})^{n_c - 1} /2.
    \label{eq:bit_flip_critical_level}
\end{equation}
where $n_c$ is given by $n_c = \lceil \frac{\alpha^2}{4} + \frac{1}{6.4}\ln \frac{2\kappa}{K} \rceil$ (we have neglected the $n$ dependence of $\kappa_{n\rightarrow n-1}$ when solving the transcendental equation $\kappa_{n\rightarrow n-1} = \chi_n^{\prime}$ to get $n_c$). For $n_c \gg 1$, $n_c!$ is approximately given by $n_c! \approx e^{n_c \ln n_c - n_c}$ using the Stirling's approximation. And by dropping the ceiling function for $n_c$ and considering the regime where $\frac{\alpha^2}{4} \gg |\frac{1}{6.4}\ln\frac{2\kappa}{K}|$, we get a simplified approximation for \cref{eq:bit_flip_critical_level}:
\begin{equation}
\begin{aligned}
\gamma_X & \approx \kappa_1 n_{\mathrm{th}}/2 \times \exp{[-(\ln 16 + 1)(\frac{\alpha^2}{4} + \frac{1}{6.4}\ln \frac{2\kappa}{K}) + \ln 4\alpha^2]} \\
& \approx \kappa_1 n_{\mathrm{th}}/2 \times \exp{(- 0.94 \alpha^2 + \ln 4\alpha^2)} \times (\frac{2\kappa}{K})^{-0.6}.
\end{aligned}
\label{eq:bit_flip_simplified}
\end{equation}
We note the approximation \cref{eq:bit_flip_simplified} works only when the suppression factor $\exp{(- 0.94 \alpha^2 + \ln 4\alpha^2)} \times (\frac{2\kappa}{K})^{-0.6}$ is smaller than 1, otherwise the bit flip rate is simply given by $\kappa_1 n_{\mathrm{th}}/2$. So overall the bit flip rate is approximately given by:
\begin{equation}
    \gamma_X \approx \kappa_1 n_{\mathrm{th}}/2 \times \min \{1, \exp{(- 0.94 \alpha^2 + \ln 4\alpha^2)} \times (\frac{2\kappa}{K})^{-0.6}\}.
    \label{eq:overall_bit_flip_rate}
\end{equation}

In \cref{fig:bit_flip_analytics} we compare the analytical expression \cref{eq:overall_bit_flip_rate} with the numerically fitted bit-flip rates. We fix the Kerr strength $K$ to be $2\pi \times 10\ \mathrm{MHZ}$, single-photon loss rate $\kappa_1$ to be $2\pi \times 1\ \mathrm{kHZ}$, thermal population $n_{\mathrm{th}}$ to be $0.1$ and evaluate the bit-flip rates for different cat size $\alpha^2$ and ideal engineered dissipation rate $\kappa_{1, \mathrm{eng}}$ (with dissipator $\kappa_{1,\mathrm{eng}}\mathcal{D}[\hat{Z}\otimes \hat{a}']$). The solid lines are theoretical bit flip rates predicted by \cref{eq:overall_bit_flip_rate} while the stars are the numerically extracted values, which are obtained by simulating the system dynamics for $50\ \mathrm{\mu s}$ and fitting the decay of $\hat{Z}$. Good agreement is shown between the theoretical and numerical values. We note that compared to the multi-mode filter model presented in the main text, the simplified model for the engineered dissipation that we consider here corresponds to a bath with spectral density that is flat over a wide range of frequency around $\omega_a - 4 K\alpha^2$ while still vanishing at $\omega_a$. We emphasize two main features of the bit-flip rate given by \cref{eq:overall_bit_flip_rate}: (1). Overall, the bit-flip rate is exponentially suppressed by $\alpha^2$, despite that for small $\kappa_{1,\mathrm{eng}}$ there is a plateau with small $\alpha^2$ before the exponential suppression. (2). Adding engineered dissipation gives an extra suppression factor $(\kappa_{1, \mathrm{eng}}/\kappa_1)^{0.6}$.


\end{document}